\documentclass[fleqn,usenatbib]{mnras}

\usepackage{newtxtext,newtxmath}

\usepackage[T1]{fontenc}
\usepackage{ae,aecompl}

\usepackage{graphicx}	
\usepackage{amsmath}	
\usepackage{amssymb}
\usepackage{ulem}
\usepackage{float}
\usepackage{epsfig}
\usepackage{subfigure}
\usepackage{epstopdf}
\usepackage{verbatim}
\usepackage{nicefrac}
\usepackage{url}
\usepackage{mathrsfs}
\usepackage{amssymb}
\usepackage{amsmath}
\usepackage{bm}
\usepackage{ulem}
\usepackage{dsfont}
\usepackage[switch,columnwise]{lineno}
\usepackage{breqn}
\usepackage{color}
\usepackage{mathtools}

\newcommand{\sigmaEight}{1.2$^{+0.9}_{-0.8}$}
\newcommand{\nDES}{196}

\newcommand{\sigmaEightJLA}{0.8$^{+1.1}_{-0.7}$}
\newcommand{\nJLA}{488}

\newcommand{\om}{\rm \Omega_{\rm m}}

\newcommand{\LCDM}{$\Lambda$\text{CDM}}

\title[Weak Lensing of SNe Ia from DES]{Weak Lensing of Type Ia Supernovae from the Dark Energy Survey}

\author[E.~Macaulay et al.]{
\parbox{\textwidth}{
\Large
E.~Macaulay$^{1,2}$ \thanks{email: \href{mailto:e.macaulay@ung.edu}{\nolinkurl{edward.macaulay@ung.edu}}},
D.~Bacon,$^{2}$
R.~C.~Nichol,$^{2}$
T.~M.~Davis,$^{3}$
J.~Elvin-Poole,$^{4,5}$
D.~Brout,$^{6,7}$
D.~Carollo,$^{6,8}$
K.~Glazebrook,$^{9}$
S.~R.~Hinton,$^{3}$
G.~F.~Lewis,$^{10}$
C.~Lidman,$^{11}$
A.~M\"oller,$^{12}$
M.~Sako,$^{7}$
D.~Scolnic,$^{13}$
M.~Smith,$^{14}$
N.~E.~Sommer,$^{11}$
B.~E.~Tucker,$^{11}$
T.~M.~C.~Abbott,$^{15}$
M.~Aguena,$^{16,17}$
J.~Annis,$^{18}$
S.~Avila,$^{19}$
E.~Bertin,$^{20,21}$
S.~Bhargava,$^{22}$
D.~Brooks,$^{23}$
D.~L.~Burke,$^{24,25}$
A.~Carnero~Rosell,$^{26}$
M.~Carrasco~Kind,$^{27,28}$
J.~Carretero,$^{29}$
F.~J.~Castander,$^{30,31}$
M.~Costanzi,$^{32,33}$
L.~N.~da Costa,$^{17,34}$
S.~Desai,$^{35}$
H.~T.~Diehl,$^{18}$
P.~Doel,$^{23}$
B.~Flaugher,$^{18}$
R.~J.~Foley,$^{36}$
J.~Garc\'ia-Bellido,$^{19}$
E.~Gaztanaga,$^{30,31}$
D.~W.~Gerdes,$^{37,38}$
D.~Gruen,$^{39,24,25}$
R.~A.~Gruendl,$^{27,28}$
J.~Gschwend,$^{17,34}$
G.~Gutierrez,$^{18}$
D.~L.~Hollowood,$^{36}$
K.~Honscheid,$^{4,5}$
D.~Huterer,$^{38}$
D.~J.~James,$^{40}$
K.~Kuehn,$^{41,42}$
N.~Kuropatkin,$^{18}$
O.~Lahav,$^{23}$
M.~Lima,$^{16,17}$
M.~A.~G.~Maia,$^{17,34}$
J.~L.~Marshall,$^{43}$
P.~Melchior,$^{44}$
F.~Menanteau,$^{27,28}$
R.~Miquel,$^{45,29}$
A.~Palmese,$^{18,46}$
A.~A.~Plazas,$^{44}$
A.~K.~Romer,$^{22}$
A.~Roodman,$^{24,25}$
E.~Sanchez,$^{26}$
V.~Scarpine,$^{18}$
M.~Schubnell,$^{38}$
S.~Serrano,$^{30,31}$
I.~Sevilla-Noarbe,$^{26}$
M.~Soares-Santos,$^{47}$
E.~Suchyta,$^{48}$
M.~E.~C.~Swanson,$^{28}$
G.~Tarle,$^{38}$
T.~N.~Varga,$^{49,50}$
A.~R.~Walker,$^{15}$
and J.~Weller$^{49,50}$
\begin{center} (DES Collaboration) \end{center}
}
\vspace{0.4cm}
\\
\parbox{\textwidth}{
Author affiliations are shown in Appendix \ref{appendix:affiliations}\\
}
}

\date{Accepted 2020 June 19. Received 2020 June 16; in original form 2020 March 25 }
\pubyear{2020}

\begin{document}
\label{firstpage}
\pagerange{\pageref{firstpage}--\pageref{lastpage}}
\maketitle

\begin{abstract}
We consider the effects of weak gravitational lensing on observations of \nDES \, spectroscopically confirmed Type Ia Supernovae (SNe Ia) from years 1 to 3 of the Dark Energy Survey (DES).  We simultaneously measure both the angular correlation function and the non-Gaussian skewness caused by weak lensing.  This approach has the advantage of being insensitive to the intrinsic dispersion of SNe Ia magnitudes.  We model the amplitude of both effects as a function of $\sigma_8$, and find $\sigma_8 =$\sigmaEight.  We also apply our method to a subsample of \nJLA \, SNe from the Joint Light-curve Analysis (JLA) (chosen to match the redshift range we use for this work), and find $\sigma_8 =$\sigmaEightJLA.  The comparable uncertainty in $\sigma_8$ between DES-SN and the larger number of SNe from JLA highlights the benefits of homogeneity of the DES-SN sample, and improvements in the calibration and data analysis.  

\end{abstract}


\section{Introduction}
\label{sec:intro}

Weak gravitational lensing (WL) is a key technique in observational cosmology \citep[e.g.,][]{2000Natur.405..143W,2010A&A...516A..63S,2012MNRAS.427..146H,2017MNRAS.465.1454H}, and also a key target of the Dark Energy Survey (DES) \citep{2005astro.ph.10346T,2016PhRvD..94b2001A,2016PhRvD..94b2002B,2018arXiv181002441O,2018PhRvD..98d3526A, 2019PhRvD..99b3508B}.  Since WL is sensitive to space-like perturbations to the background cosmological metric, WL observations are a crucial cosmological test of theories of modified gravity \citep[e.g., ][]{2005PhRvD..71h3512S,2007A&A...463..405S,2008PhLB..665..325T,2008PhRvD..78d3002S}.

The most established method of WL is the measurement of coherent distortions in the shapes of galaxies, particularly the shear \citep[e.g.,][]{2017MNRAS.471.1259J,2018MNRAS.481.1149Z,2018PhRvD..98d3528T}.  WL has also been detected in observations of the Cosmic Microwave Backround (CMB);  \citep[e.g.,][]{2018arXiv181002441O, 2018arXiv180706210P}.  

WL is also expected to affect the magnitudes of standard candles.  Magnification due to WL has been detected in quasars  \citep[e.g.,][]{2005ApJ...633..589S,2011ApJ...732...64B}, and also studied with Type Ia Supernovae (SNe Ia) \citep[e.g.,][]{1997ApJ...475L..81W,2000A&A...356..771V,2004ApJ...606..654W,2005JCAP...03..005W,2006PhRvD..74f3515D}.  We note that weak gravitational lensing is also expected to affect gravitational-wave standard sirens, e.g. \cite{2011MNRAS.411....9S} , \cite{2017NatCo...8.1148L}.

Over-dense lines of sight enhance the focus of light rays, causing a magnification in the observed brightness, whereas under-dense lines of sight lead to a de-magnification.  \cite{2014ApJ...780...24S} estimated the expected lensing magnification of SNe in the Joint Light-curve Analysis (JLA) SN sample \citep{2014A&A...568A..22B}.  The expected lensing magnification for each SN was calculated by integrating the line-of-sight densities of galaxies observed by the Sloan Digital Sky Survey (SDSS) Baryon Oscilation Spectroscopic Survey (BOSS);  \citep{2013AJ....145...10D}. A correlation between the magnitude residuals and the expected lensing signal was found at the 1.7 $\sigma$ confidence level.

Since most lines of sight in the Universe are under-dense, and only rare lines of sight are over-dense, WL causes a characteristic skewness in the distribution of magnitude residuals of standard candles.  The amount of lensing skewness depends on the contrast between voids and over-densities, and as such depends on $\Omega_m$ and $\sigma_8$.

In order to provide fast calculations of one-dimensional lensing statistics, \cite{2009PhRvD..80l3020K} developed the \texttt{turboGL} code, which was developed further in \cite{2011PhRvD..83b3009K}.  These lensing statistics were the basis for a series of papers developing a method to estimate the likelihood of the lensing distribution of a supernova sample (MeMo, `the Method of the Moments'); \citep{2013PhRvD..88f3004M,2014PhRvD..89b3009Q}.

Instead of enforcing a fiducial distribution for the SN magnitude residuals, MeMo is based on empirical fitting functions for the first, second, third, and fourth moments of the residuals of SNe distance measurements, which have been calibrated to N-body simulations as a function of redshift and parameters $ \om $, $\sigma_8$, and the intrinsic dispersion in SN magnitudes, $\sigma_{\rm{int}}$.  In \cite{2014MNRAS.443L...6C}, MeMo was applied to the JLA sample, finding $\sigma_8 = 0.84^{+0.28}_{-0.65}  $ (at the one-standard deviation confidence level).

The effect of WL on SN magnitudes shares some similarities with the effects of Peculiar Velocities (PV), although there are also some significant contrasts \citep{2007PhRvL..99h1301G,2007ApJ...661L.123N,2011ApJ...741...67D,2016PDU....13...66C,2019arXiv190500746G}.  Both effects are sensitive to cosmological density fluctuations.  However, PVs are sensitive to large scale ($\sim$ Gpc) density fluctuations, whereas WL is most sensitive to far smaller ($<$ Mpc) scales. 

Whereas WL directly affects the observed magnitudes, PVs mainly\footnote{There is a direct effect on the observed magnitude from the `$(1+z)$' term in the luminosity distance, although this is not the dominant effect.} affect the redshift; the apparent effect on the magnitude residual is due to comparing the theoretical magnitude at the PV-affected redshift, instead of the `true' cosmological redshift.

While both effects are similar in magnitude at $z \sim 0.2$, PVs dominate at lower redshifts, while WL becomes dominant at higher redshifts.  Over the redshift range 0.1 $< z <$ 0.3, the two effects are similar in magnitude \citep{2017MNRAS.467..259M}.

Since many SNe in the JLA catalogue are within this redshift range, in \cite{2017MNRAS.467..259M} we built on the WL MeMo method to include the effects of PVs on the moments of magnitude residuals.  Modelling both PVs and WL allows for a simultaneous test of the consistency of Newtonian and lensing metric perturbations, which is a key prediction of the \LCDM $\,$ model \citep[e.g.,][]{2004PhRvD..70d3009H,2006PhRvD..74b3512K,2007PhRvL..98l1301K,2007PhRvL..99n1302Z,2008PhRvD..78b4015B,2013MNRAS.429.2249S}. 

However, in this paper, we consider only the effects of WL, since the effect of PVs is sub-dominant, due to the absence of low redshift ($z<0.2$) SNe in the DES-SN sample.  

In addition to the effect on the moments of residuals, WL also causes correlations in the magnitudes.  \cite{2017MNRAS.465.2862S} modelled the signal to noise ratio of the lensing correlation function, and forecast the signal-to-noise ratio of a WL detection by LSST.  \cite{2017MNRAS.465.2862S} showed that for a SN survey the size of DES-SN (\nDES \, SN), the effects of WL on the correlation function would not be detectable.  Nevertheless, measurements of the correlation function (however noisy), may still place upper limits on the effect of WL on the magnitude residuals.

Since the correlation function is (in principle) independent of the skewness, we can combine both observations.  However, we may expect covariance between the two different types of measurements, since both observations are drawn from the same sample of data.  In order to account for any covariance, we build on the bootstrap resampling method used in \cite{2017MNRAS.467..259M} to allow for two different types of observable.  We describe this method further, as well as the relevant theory and details of the observations, in Section \ref{sec:data}.  We discuss our results and conclusions in Section \ref{sec:Discussion}.

\section{Data \& Methodology }
\label{sec:data}

We study \nDES \, new spectroscopically confirmed Type Ia SNe from the Dark Energy Survey (DES).  The observation and reduction of the sample is described in a series of papers in \cite{ 2018arXiv181102377B, 2018arXiv181102378B, 2018arXiv181109565D,2018arXiv181102380L,2019MNRAS.tmp..472K}.  The cosmological analysis of the sample is described in \cite{2019ApJ...872L..30A}, who found a value of $\om =  0.331 \pm 0.038$ for a flat \LCDM \, cosmology.  This is the fiducial cosmology we assume throughout this work.  

\begin{figure}
\begin{center}
 \includegraphics[width=8cm]{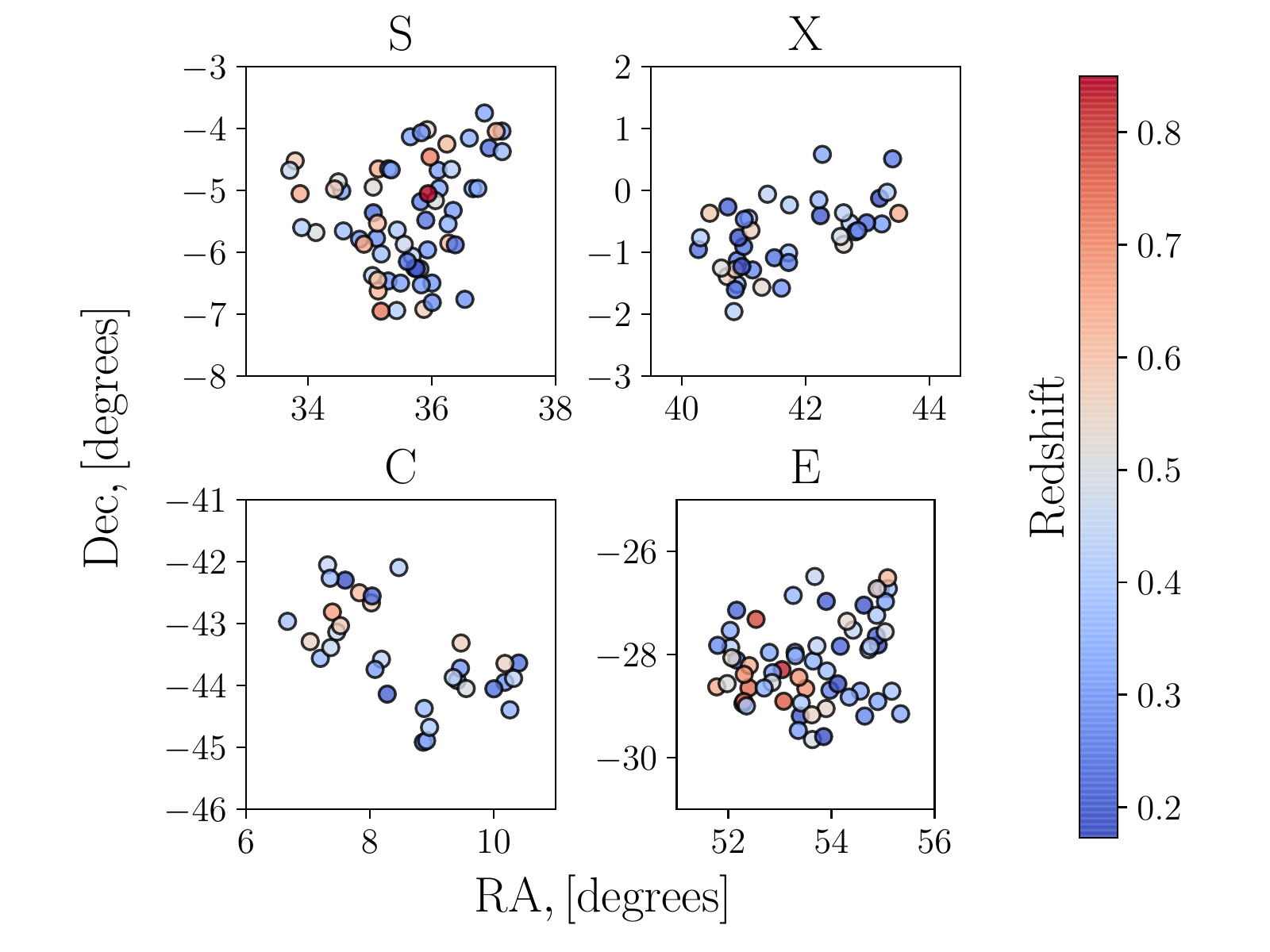}
\caption{The angular coordinates of the four DES supernova fields.  Due to the large angular separations between fields, we do not consider correlations between different fields.  The points have been colour-coded by redshift; blue for lowest $z$, and red for highest $z$.}
   \label{fig:DES_SN_RA_dec}
   \end{center}
\end{figure}

We illustrate the angular coordinates of the SNe in Figure \ref{fig:DES_SN_RA_dec}.

The SNe are located in the four DES SNe fields, chosen to overlap with other well-studied fields: `S' (SDSS Stripe 82), `X' (XMM-LSS), `C' (Chandra Deep Field South), `E' (Elais-S1).

We also compare our results from DES-SN to SN from the JLA sample \citep{2014A&A...568A..22B}.  The full JLA sample consists of 740 SNe spanning a redshift range up to $z=1.4$.  We consider only a sub-sample of \nJLA \, in the redshift range of our lensing model, $0.2<z<0.5$.
 
 For both DES-SN and JLA, we consider the distance-redshift relation fixed (using the best-fit cosmology for each survey), since the $\sim10$\% uncertainty in the distance-redshift relation is small compared to the $\sim100 \%$ uncertainty in the lensing signal.  

We consider two different effects caused by WL on SN magnitudes: the correlation in the residuals (due to the fact that the lensing signal will be similar for close lines of sight), and the skewness (caused by the asymmetry of over and under-densities of dark matter).  We anticipate that with the final SN sample from DES, the correlation function will be of particular interest, due to the high angular density of the SN in the small area (30 deg$^2$) of the SN survey.  We base our treatment of the skewness on the MeMo approach \citep{2013PhRvD..88f3004M,2014PhRvD..89b3009Q,2014MNRAS.443L...6C}.

\subsection{MeMo}

MeMo is based on minimizing the $\chi^2$ given by
\begin{equation}
\label{eq:MeMo}
\chi^2 = \sum_{i} \bmath{ \Delta \mu}_i \textbfss{C}_{i} ^{-1} \bmath{ \Delta \mu}_i^{\intercal},
\end{equation}
where $\Delta \bmath{\mu}_{i}$ is a vector given by
\begin{equation}
\label{eq:DeltaMu}
\bmath{ \Delta \mu}_i = \bmath{\mu}^{\rm{Obs}}_{i} - \bmath{\mu}^{\rm{MeMo}}_{i}  .
\end{equation}
Here, $\bmath{\mu}^{\rm{Obs}}_{i}$ is a vector of the observed first, second, third, and fourth moments of the distance moduli of the SN within the $i^{\rm{th}}$ redshift bin, and $\textbfss{C}_{i}$ is the corresponding data covariance matrix of this vector.  $\bmath{\mu}^{\rm{MeMo}}_{i}$ is the theoretical expectation from the MeMo fitting functions.    

The theoretical expectations for the moments are calculated from fitting functions, in terms of $z$, $\Omega_m$, and $\sigma_8$.  The fitting functions for the second, third, and fourth moments are shown in full in Equations 6,7, and 8 in \cite{2013PhRvD..88f3004M}.  We note that the even moments both depend on the intrinsic dispersion of SN magnitudes, whereas -- under the assumption that the intrinsic dispersion is Gaussian --  the third moment depends only on the cosmological parameters.

The functions are based on the \texttt{turboGL} code, and tested against simulation results by \cite{2008MNRAS.386.1845H} and \cite{2011ApJ...742...15T}.  This approach has the advantage of computational efficiency in evaluating the moments, however, it is limited to the range of parameters used to evaluate the moments, and the cosmological model used in the simulations (flat \LCDM).

\cite{2018MNRAS.478.1305C} used simulations to study the effect of baryonic physics on 1-point lensing statistics, finding that the effect of baryons could double the probability of highly lensed objects.

In \cite{2014MNRAS.443L...6C},  $\textbfss{C}_{i}$ was calculated analytically based on observations of higher moments of the SN residuals.  However, in \cite{2017MNRAS.467..259M}, we found that the necessity of the analytical estimation of the data covariance matrix on measurements of high moments (up to the $8^{\rm{th}}$) could lead to biases in the estimation of $\textbfss{C}_{i}$.  Instead, we adopted a bootstrap re-sampling method, based only on the directly observed moments (i.e., up to the fourth).  This bootstrap resampling method naturally allows for the covariance to be estimated between different types of observables, which is the approach we take here to include covariance between observations of the correlation function and the skewness of the SN.

\subsection{Correlations}

The correlation function of magnitude residuals due to weak lensing is given by 
\begin{equation}
\label{eq:corr_func}
\left< \Delta \mu_i, \Delta \mu_j \right> = \left( \frac{5}{\ln 10} \right)^2 \left( \frac{1}{2 \pi } \right)  \int^{ \infty }_{0}  \ell  P_{\kappa}( \ell ) J_0(\ell \Theta) d \ell,
\end{equation}
where $J_0$ is the Bessel function of the zeroth-kind \citep{2001PhR...340..291B}.  $P_{\kappa}( \ell )$ is the angular power spectrum of the lensing convergence, $\kappa$, which is related to the matter power spectrum $P_{\delta}$ by
\begin{equation}
\label{eq:corr_func}
P_{\kappa}( \ell ) = \frac{9 H_0^4 \om^2   }{4 c^4}  \int^{ \chi_H }_{0}  d \chi \frac{W^2( \chi )}{a^2( \chi )} P_{\delta} \left( \frac{\ell}{\chi},\chi \right),
\end{equation}
where $W$ is the redshift distribution of the sources, and $a$ is the scale factor, both as a function of comoving distance, $\chi$.  $H_0$ is the Hubble parameter, $\om$ is the total matter density, and $c$ is the speed of light.  We calculate $P_{\kappa}( \ell )$ with \texttt{CAMBSources}  \citep{2011PhRvD..84d3516C}, based on the redshift distribution of the SNe.  In order to quantify the SNe redshift distribution, we use Kernel Density Estimation (KDE) based on the redshifts of the SNe.

In this work, we focus on small scale ($\Theta<30$ arcmin) angular correlations.  This is because the strongest correlation is expected at $\Theta<10$ arcmin scales (as can be seen in Figure \ref{fig:correlation_sim}).

\subsection{Combined Observations Likelihood}

In this paper, we simultaneously fit for the correlation and skewness caused by gravitational lensing, by minimising the $\chi^2 $  given by
\begin{equation}
\label{eq:combined}
\chi^2 =  \bmath{ \Delta_\mu} \, \textbfss{C} ^{-1} \, \bmath{ \Delta_\mu}^{\intercal},
\end{equation}
where (as with MeMo), $ \bmath{ \Delta_\mu} $ is the difference between theory and observation vectors.  We note that since our estimate of the data covariance matrix does not depend on $\sigma_8$,  minimising this $\chi^2 $ is equivalent to likelihood maximisation (for a fixed data covariance matrix).  The vector $ \bmath{\mu} $ is constructed by concatenating the vectors for the two different types of observations:
\begin{equation}
\label{eq:data_vector}
\bmath{ \mu} = \left[\bmath{\mu}_{ij}    ,  \bmath{\mu}_{3}  \right] ,
\end{equation}
where $\bmath{\mu}_{ij} $ is the correlation vector (as a function of angular separation), and $\bmath{\mu}_{3}$ is the third moment vector (as a function of redshift).  We use four bins of angular separation, and five redshift bins.  

We choose the redshift binning to match the redshift bins in \cite{2019ApJ...872L..30A} (within the redshift range used here).  We have verified that the method is stable when changing the number of redshift or angular bins.  This stability is a aided by the bootstrap-resampling method used to estimate the covariance matrix: increasing the number of bins naturally increases the correlations in the off-diagonal elements of the covariance matrix.

We do not include the second and fourth moments, since the unique lensing skewness is captured by the third moment, and the even moments are sensitive to the model of the SN intrinsic dispersion. The implicit assumption of this choice is that the intrinsic dispersion of the magnitudes has zero skewness.  We use \texttt{emcee} \citep{2013PASP..125..306F} to probe our likelihood.

\subsection{Covariance Matrix}

\begin{figure}
\begin{center}
 \includegraphics[width=8cm]{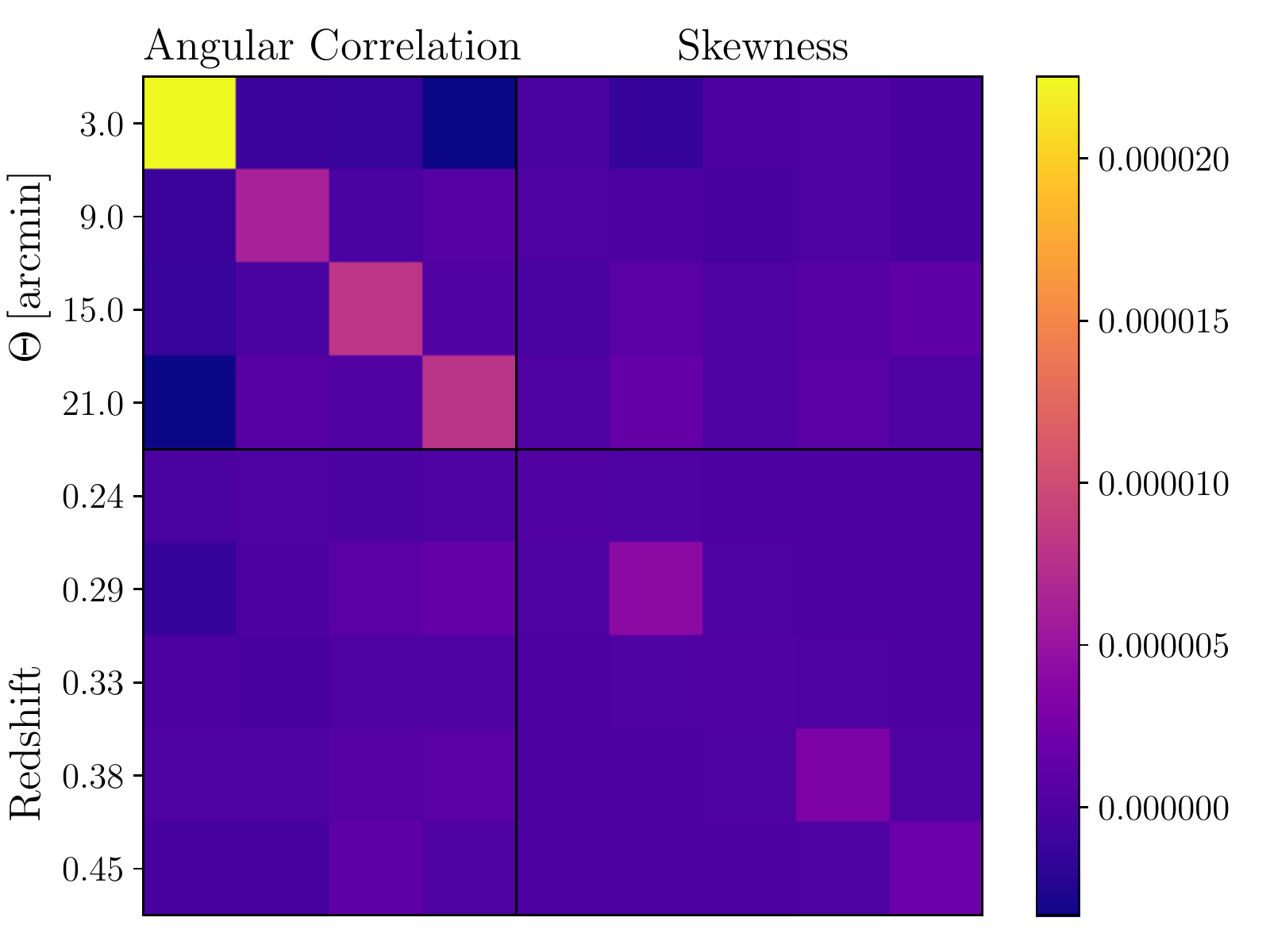}
\caption{Illustrating the data covariance matrix.  The upper-left corner of the matrix is from the angular correlation function, and the lower-right corner is from the skewness measurements.  The bootstrap resampling method we use to estimate the covariance matrix allows us to estimate the cross-correlations between these two types of observable, shown in the off diagonal blocks.}
   \label{fig:combined_CM}
\end{center}
\end{figure}

In a similar manner to \cite{2017MNRAS.467..259M}, we use a bootstrap re-sampling method to estimate the data covariance matrix.  To estimate $\textbfss{C}$, we randomly sample (with replacement) from the genuine sample of SNe Ia.  For each random sample, we calculate a combined data vector, consisting of the correlation and skewness vectors.  We repeat this process of sampling to generate an ensemble of measurements, from which we can directly calculate an estimate of the covariance matrix.  The advantage of this approach is that we can naturally estimate the covariance between the two different types of observable (the correlation and the skewness).  The data covariance matrix is illustrated in Figure \ref{fig:combined_CM}.

\subsection{Simulations}

In order to verify our methodology against a fiducial cosmology, we generate realizations of DES-SN by drawing samples from the \texttt{MICE} simulation \citep{2015MNRAS.447.1319F} with selection functions to match the genuine observations.  We note that we use the simulations only as a test of the methodology; they do not directly contribute to the analysis of the genuine data (e.g., we do not use the simulations to estimate the data covariance matrix, etc.).

We start with a 3,000 square-degree light-cone from \texttt{MICE}, comprising over one million simulated galaxies.  We then perform an angular selection cut, to match the 30 square-degree area of the supernova fields in DES.  We match the footprints of the DES SNe fields, e.g., three patches for the C and X fields, and two for the E and S fields, so that the angular footprint matches the genuine DES fields. Over ten-thousand simulated galaxies pass this angular selection cut. 

We next sub-sample to match the number and redshift distribution of DES-SN.  We approximate the redshift probability distribution function of DES-SN with a 1D KDE in redshift.  To generate each simulated realization, we then draw \nDES \, galaxies from the super-sample of ten-thousand, with a probability ascribed to each galaxy given by the 1D KDE.  Each realization thus matches the angular and redshift distribution of the genuine sample.

For each of these galaxies, we take the value of the lensing convergence, $\kappa$, as estimated by the MICE simulation, and calculate the magnitude residual $\Delta_{\mu}$ due to $\kappa$, using: 
\begin{equation}
\label{eq:mag_kappa}
\Delta_{\mu} = 5  \log_{10}( 1 - \kappa )
\end{equation}
\citep{2001PhR...340..291B}.  To include the effect of the intrinsic dispersion for a SN, we then add to the lensed-only residual $\Delta_{\mu}$ a noise term, sampled from a Gaussian distribution (with standard deviation of 0.1 mag).  This $\Delta_{\mu}$ thus approximates a SN magnitude residual, including the intrinsic dispersion and lensing magnification.

\begin{figure}
\begin{center}

\includegraphics[width=8cm]{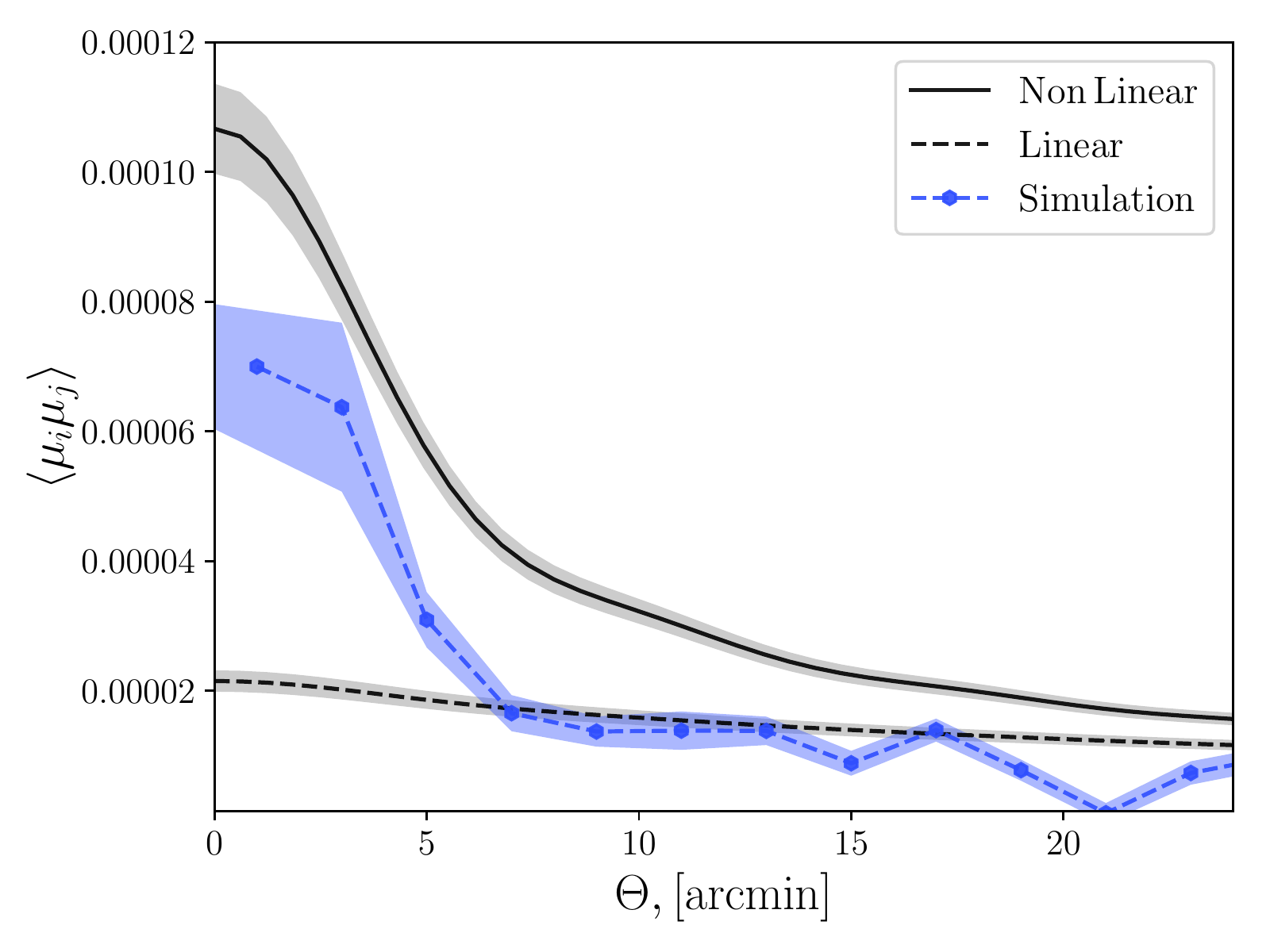}
\caption{Measurements of the lensing correlation function measured from the \texttt{MICE} simulation to theoretical expectations from \texttt{CAMBSources}.  The solid black line is the expected correlation function assuming a non-linear power spectrum, and the dashed black line assumes a linear power spectrum.  The grey shaded regions illustrate the one standard-deviation regions of these functions, due to small differences in the redshift distributions of each realization.  The blue hexagonal points illustrate the average measured lensing correlation function from the \texttt{MICE} simulation.}
   \label{fig:correlation_sim}

    \includegraphics[width=8cm]{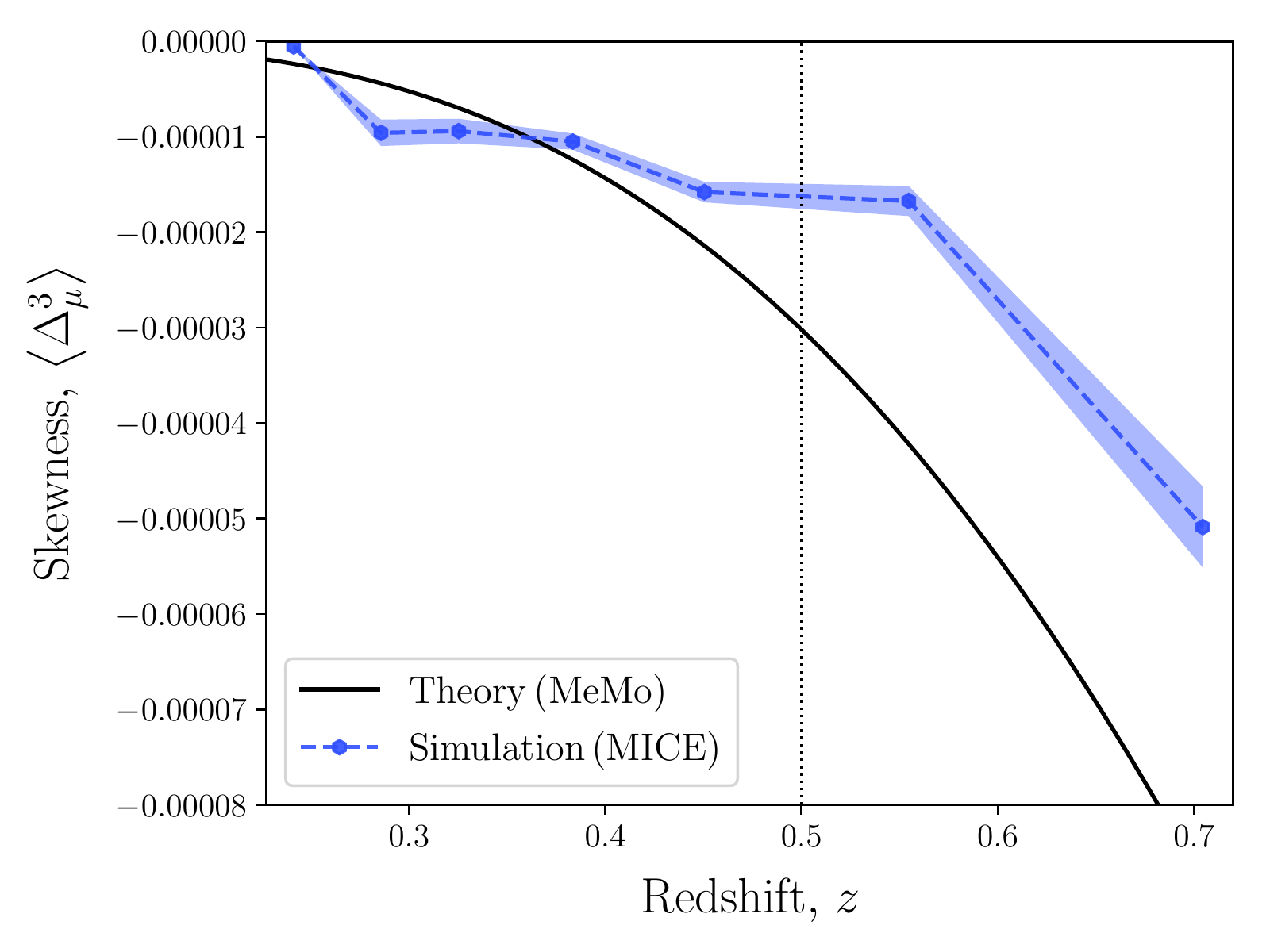}
\caption{In this figure we compare measurements of the lensing skewness from the \texttt{MICE} simulation to the corresponding theory from the MeMo fitting function.  We note that \texttt{MICE} appears to under-predict the skewness, compared to the MeMo fitting function for $z>0.5$. 
}
   \label{fig:skewness_sim}

\end{center}
\end{figure}

In Figure \ref{fig:correlation_sim} we illustrate the measured correlation function from the \texttt{MICE} simulations.  Since the signal from each realization is low, we measure the correlation function for 10,000 realizations, and show the average.  We compare the simulated measurements to the expected theoretical linear and non-linear correlation function for the fiducial cosmology used in the \texttt{MICE} simulation.  

The theoretical correlation function depends on the redshift distribution of the sources, which varies between realizations due to sampling noise.  We re-calculate the theory for each redshift distribution, and illustrate the range in Figure \ref{fig:correlation_sim}.

We can see in Figure \ref{fig:correlation_sim} that \texttt{MICE} most closely matches the linear theory for angular scales of $\Theta > 6$ arcmin.  For smaller angular scales, \texttt{MICE} is closer to the non-linear theory.  This may be due to the resolution of the \texttt{MICE} simulation. 

MICE is based on a Gadget-2 simulation \citep{2005MNRAS.364.1105S} with a box width of 3,072 $h^{-1}$ Mpc, and a softening length of 50 $h^{-1}$ kpc.  In order to construct the simulated light-cones \cite{2015MNRAS.447.1319F} sample the simulation into 265 radially concentric `onion shells' over the redshift range $0<z<1.4$, each corresponding to a time of approximately 35 megayears per shell.  To calculate the lensing convergence maps, \cite{2015MNRAS.447.1319F} pixelate each shell using a \texttt{Healpix} angular tessellation with a \texttt{Healpix} resolution of $N_{\rm{side}}$=4,096.  This tessellation corresponds to an effective angular resolution of 0.85 arcmin.  This resolution remains constant in redshift, since each shell is tessellated with the same $N_{\rm{side}}$=4,096.  This limit may be too coarse to fully resolve the rare, dense structures predicted by the theoretical non-linear correlation function.

Even though the resolution of the MICE simulation is 0.85 arcmin, we believe there are at least two effects that may introduce suppression of the correlation function out to angular scales larger than the resolution limit.  Firstly, feedback from smaller scales will not be able to affect structure formation at larger scales.  In other words, the simulation will not resolve additional gravitational accretion from dense, sub-resolution structures.  Secondly, even if structure at scales greater than 0.85 arcmin were perfectly resolved, we would still expect the correlation function to be under-estimated for these scales, if the density of structure at sub-resolution scales remains under-estimated, since the correlation function depends on the product of the densities across a range of angular scales.

We note that \cite{2015MNRAS.447.1319F} also find that for angular scales approaching 1 arcmin ($\ell \sim 10,000$), the convergence power spectrum underestimates the non-linear theoretical expectation by $\sim 30$ \%.  This divergence can be seen in Figure 2 of \cite{2015MNRAS.447.1319F}, particularly for $\ell$ close to 10,000. 

We note that since the correlation function is an integral of the power spectrum across a range of scales, if the power spectrum is systematically underestimated across the range of integration, a given scale in the correlation function will be underestimated by a greater amount than the direct value of the power spectrum at the direct scale corresponding to the scale in the correlation function.  We believe this is 
consistent with the results in Figure \ref{fig:correlation_sim} for the angular correlation function from the MICE simulation.   

We believe there is a clear opportunity for future work to improve simulation of the small-scale lensing convergence power spectrum.  However, we emphasise that since the current data places only large upper-limits on the correlation function, this is not a limiting factor for this analysis.

When fitting for the correlation function, we assume a fiducial form for the function (given by the background cosmology and redshift distribution of the sources), and scale the amplitude according to the value of $\sigma_8$.  To fit for the genuine data, we use \texttt{CAMBSources} to calculate the matter power spectrum and lensing window function, with HALOFIT \citep{Smith_2003} to calculate the non-linear power spectrum.

However, we can see in Figure \ref{fig:correlation_sim} that this will over-predict the lensing signal in the simulated data.  As such, when we fit for simulated catalogues, instead of using either the linear or non-linear correlation functions, we take the ensemble average correlation function  in Figure \ref{fig:correlation_sim} as the fiducial.

Although we parameterize the amplitude of the density perturbations with $\sigma_8$, we note that the physical scale of the density perturbations which affect SN lensing are at smaller physical scales (or higher $k$ values, in the case of the power spectrum) than the 8 Mpc scale directly associated with $\sigma_8$.  Our assumption is that once we have set the amplitude of the density perturbations (via $\sigma_8$), we can extrapolate the scale of density fluctuations to smaller scales.  Our use of $\sigma_8$ as a parameter is to set the relative scale of the density fluctuations, and not necessarily as a direct probe of structure at 8 Mpc scales.

We also test the skewness from the \texttt{MICE} simulation against the theoretical expectation from MeMo, shown in Figure \ref{fig:skewness_sim}.  We find that  for $z>0.5$ the \texttt{MICE} simulation under-predicts the effect of lensing moments compared to the MeMo fitting function.

 We believe this is consistent with the under-prediction of correlations at small angular scales seen in Figure \ref{fig:correlation_sim}.  Both the high $z$ skewness and the small $\Theta$ correlation function rely on the simulation of small, rare, dense lines of sight, which demand high simulation resolution, and accurate modelling of the complicated baryonic physics which affects these small scales.  

\cite{2018MNRAS.478.1305C} found that introducing baryons has a significant effect on the lensing probability density function (although at scales that would not be resolved by the ~0.85 arcmin resolution of MICE).

As we consider the lensing signal for lines-of-sight at increasingly high redshift, the chance of finding a rare, dense structure (causing high magnification) will increase.  If these structures are not resolved in the simulation, we would expect the measured skewness to be less than expectations.  As such, we do not fit for any moments for SNe with $z>0.5$ for the real or simulated data sets.  Although this redshift cut reduces our sensitivity to the most highly lensed supernovae, the sparseness of the DES-SN sample at these redshifts also makes robust measurements of the skewness difficult.

\section{Discussion of Results \& Conclusions}
\label{sec:Discussion}

\begin{center}
\begin{table*}

\begin{tabular}{c c c r }
 Reference & Sample  & Method  & $\sigma_8$ Value   \\
\hline
\cite{2014MNRAS.443L...6C} & JLA  & Variance+Skewness+Kurtosis  & $0.8^{+0.3}_{-0.7}  $   \\
\cite{2017MNRAS.467..259M} & JLA  & Variance+Skewness+Kurtosis  &$1.6^{+0.5}_{-1.0}  $   \\
 This Work & JLA  & Skewness+Correlation  & \sigmaEightJLA   \\
 This Work & DES-SN  & Skewness+Correlation  & \sigmaEight   \\

\end{tabular}
\caption{A summary of $\sigma_8$ measurements with SN-WL. We note that with the same method, DES-SN achieves smaller uncertainty than JLA, despite being a smaller sample.  We also note in a comparison of methods (both with JLA), that the method in this work achieves similar uncertainty to that of \protect\cite{2017MNRAS.467..259M}, without the imposition of a constant, Gaussian model for the SN intrinsic dispersion.  Although the smallest uncertainties in this comparison are from \protect\cite{2014MNRAS.443L...6C}, we note that this work makes the least conservative assumptions here as to the measurements of the moments, and the estimation of the data-covariance matrix \protect\citep[as discussed further in ][]{2017MNRAS.467..259M}.}  
\label{tab:results}
\end{table*}
\end{center}

\begin{figure}
\begin{center}
 \includegraphics[width=8cm]{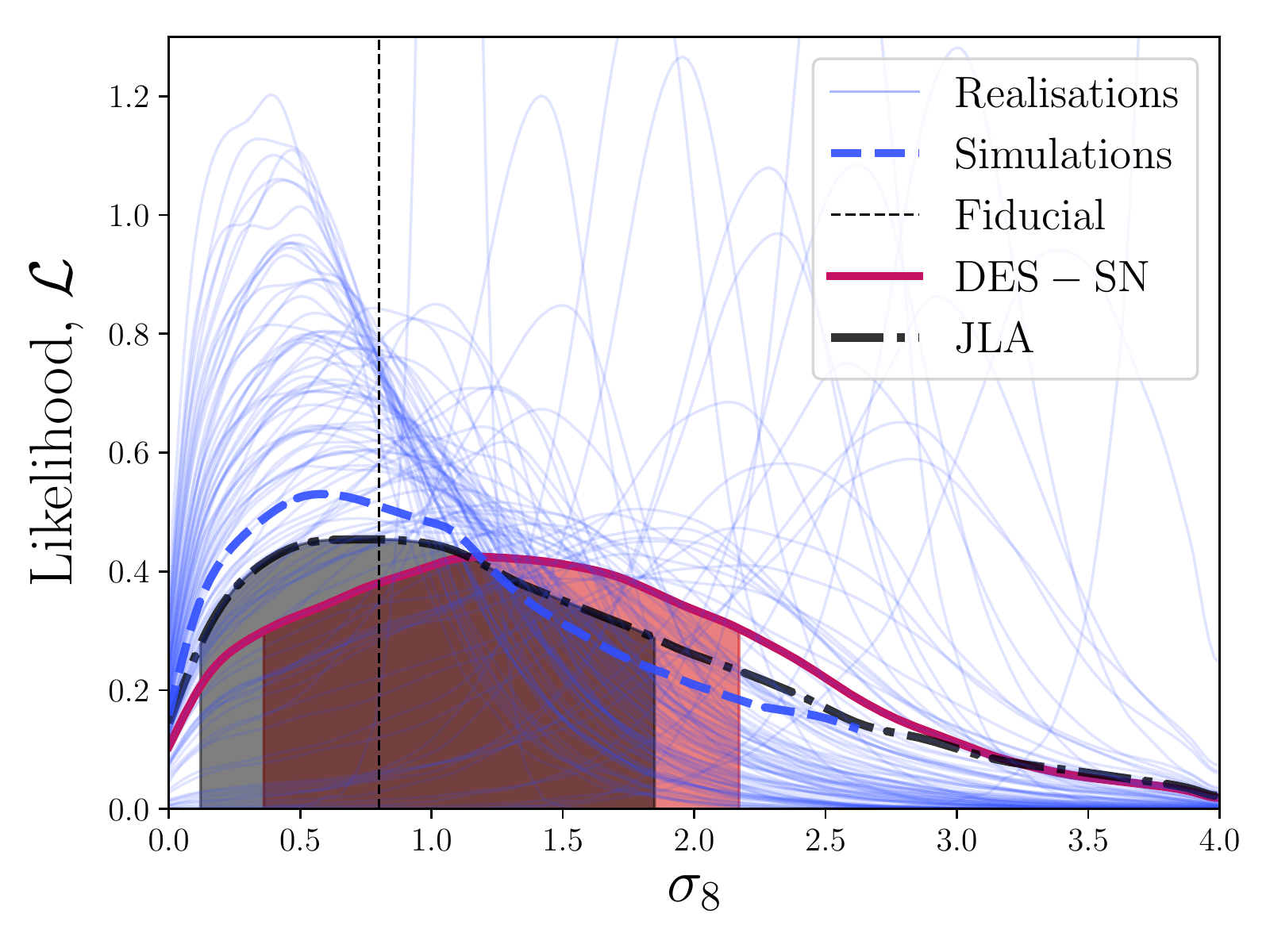}
\caption{Constraints on $\sigma_8$ from the genuine DES-SN sample, and also simulated realizations of the sample.  Individual fits to simulated realizations are shown in thin blue lines, and the average of these fits is shown with a thick, dashed blue line.  The input value of $\sigma_8$ used in the simulation is shown with the thin, black vertical dashed line. The thick red line illustrates the constraints from the genuine sample, with the one-standard deviation uncertainty range shown with the shaded region. }
   \label{fig:likelihood_plot}
\end{center}
\end{figure}

The results of the fits to 100 simulated data-sets are shown in Figure \ref{fig:likelihood_plot}.  We plot the likelihood for each simulation, and also illustrate the average of the likelihoods, in order to verify that the results are consistent with the input of the simulation.  

We note that the simulated realizations tend to cluster in two groups: a large fraction which recover a lower value of $\sigma_8$ than the simulation input, and a small fraction which recover a higher value.  We believe that this is because of the non-Gaussian lensing probability distribution function, combined with the limited size of each realization.  Due to the low likelihood of sampling from the high magnification tail, many of the realizations of \nDES \, SNe will not have any samples drawn with high magnification, which will lead to low values of $\sigma_8$.  Conversely, a minority of the samples will have several magnified SNe, leading to an over-estimated value of $\sigma_8$.  We believe this causes the distribution of simulated realisations in Figure \ref{fig:likelihood_plot}; with many realisations under-estimating $\sigma_8$, and a smaller sample over-estimating $\sigma_8$.   However, we note that the ensemble average of the realizations is not biased.

\begin{figure}
\begin{center}

 \includegraphics[width=8cm]{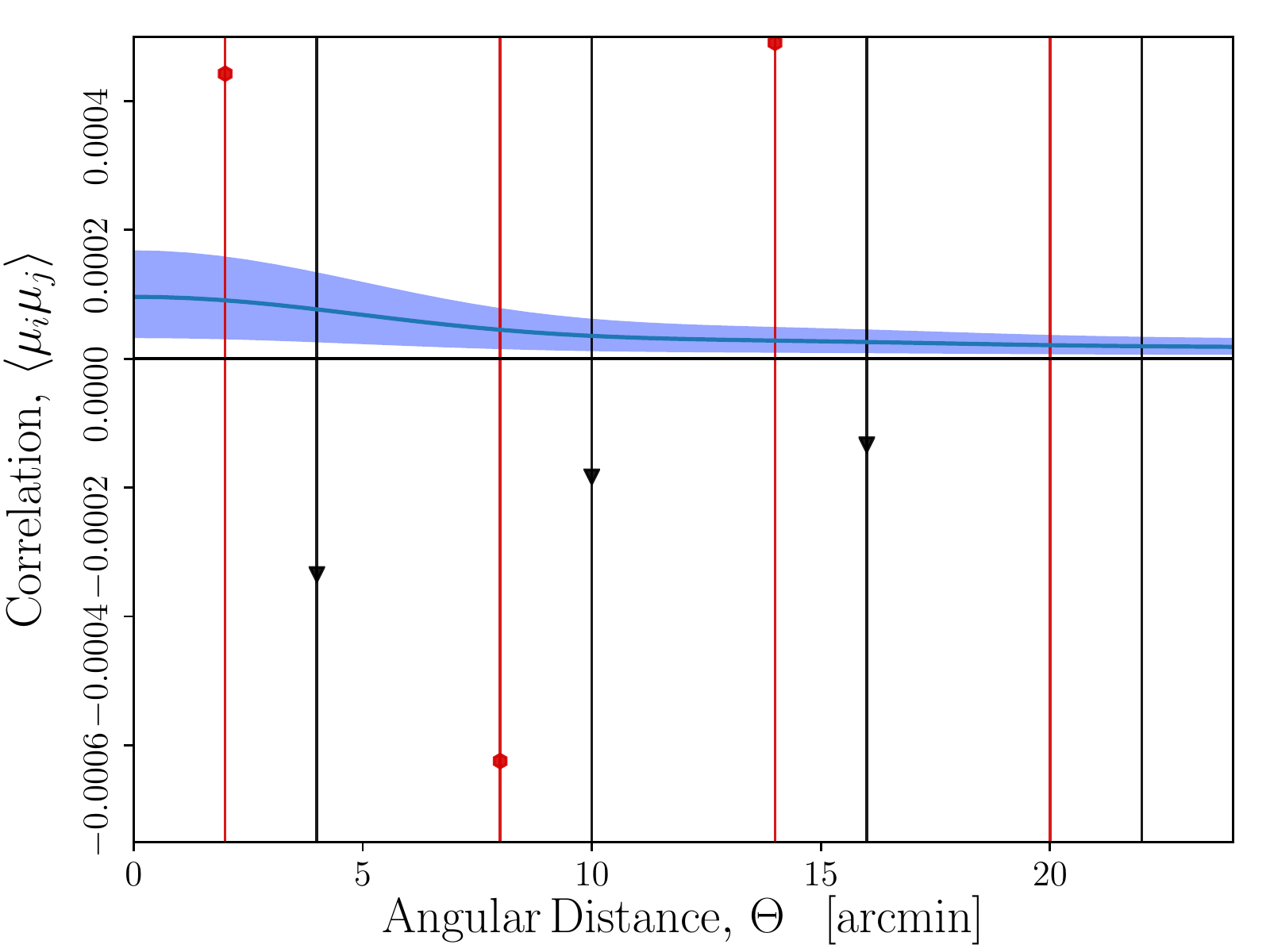}
\caption{The observed angular correlation function of the JLA and DES-SN samples.  The fiducial angular correlation function is illustrated with the blue line, and the allowed range of $\sigma_8$ from DES-SN is shown with the shaded region.  We note that the measured points are highly correlated, and in some cases we can place only upper-limits on the angular correlation function.}
   \label{fig:plot_correlation}

 \includegraphics[width=8cm]{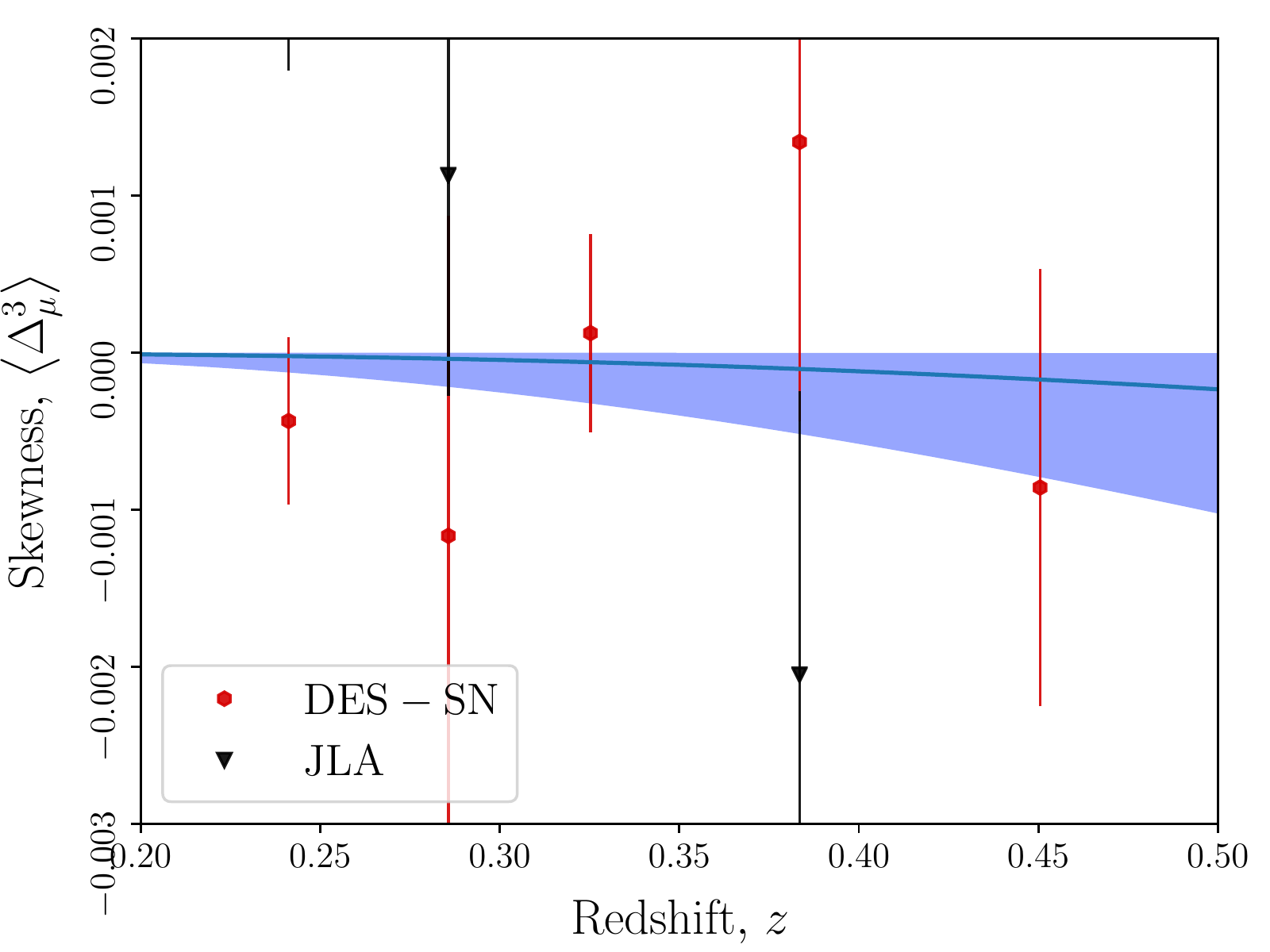}
\caption{The measured skewness of the JLA and DES-SN samples.  The blue shaded region illustrates the range of lensing skewness given the range of the best-fit $\sigma_8$ from the DES-SN sample ($\sigma_8$=\sigmaEight). }
   \label{fig:plot_skewness}
   
\end{center}
\end{figure}

We also show in Figure \ref{fig:likelihood_plot} the equivalent fit to the genuine DES-SN data.  We find $\sigma_8 =$\sigmaEight, suggesting a possibility for WL at the $\sim 1.3 \sigma$ level.  We note that the uncertainty of our measurement is consistent with the uncertainties from the simulated data.  In Figures \ref{fig:plot_correlation} and \ref{fig:plot_skewness} we plot the observed angular correlation function and skewness for DES-SN.  

 Albeit within large ($\sim100$ \%) uncertainties, we note that our value is consistent with the value of  $\sigma_8 = 0.8120 \pm 0.0073$ derived by \cite{2018arXiv180706209P} from measurements of the CMB, assuming a \LCDM $\,$ model. The values are summarised in Table \ref{tab:results}.

Perhaps of more significance than the best-fit value of $\sigma_8$, we emphasize the lower limit of $0.4$.  We note that this lower limit is not sensitive to the value of the intrinsic dispersion of the SNe Ia, as with, e.g., \cite{2014MNRAS.443L...6C} and \cite{2017MNRAS.467..259M}.  We assume only that the intrinsic dispersion is uncorrelated (i.e., genuinely intrinsic to each SN), and does not introduce additional skewness.  Beyond these assumptions, the model requires no further assumption as to the intrinsic dispersion of the SNe (such as redshift independence). 

Although far from a detection of WL in SNe, this lower limit on $\sigma_8$ suggests some possibility for WL, consistent with the statistical limitations of the size of the sample \citep[e.g.][]{2017MNRAS.465.2862S}.  We note that the limits on WL is consistent with the lensing results found by \cite{2014ApJ...780...24S}, who found limits on the lensing signal with a significance of 1.7$\sigma$.  The value of $\sigma_8$ and significance of the lensing signal is also consistent with the values from \cite{2017MNRAS.467..259M}.

We repeat our measurement with the JLA sample.  As with DES-SN, we restrict the redshift range to $z>0.17$, to minimize sensitivity to peculiar velocities.  We also restrict the redshift range to $z<0.5$ (again, as with DES-SN), since we believe that the systematic uncertainties in theoretical modelling and simulation of extreme-small scale weak lensing at high redshift is currently insufficient, even given the large statistical uncertainties of WL in current SNe surveys.  After these redshift cuts, we have \nJLA  $\,$ SNe from the JLA sample, and find $\sigma_8 =$ \sigmaEightJLA.

We note that the range of uncertainty $\sigma_8$ from JLA of 1.8 is larger than the range from DES-SN, of 1.7, despite the larger number of \nJLA $\,$ SNe in the JLA sub-sample than the \nDES $\,$ from DES.  We believe this is due to the homogeneity of the DES-SN sample, and improvements in photometric calibration. 

We note that our values with both DES-SN and JLA are consistent with the value from  \cite{2014MNRAS.443L...6C}, studying JLA, who found $\sigma_8 = 0.84^{+0.28}_{-0.65}  $.   Comparing directly to our value with JLA of $\sigma_8 =$ \sigmaEightJLA, we note that although our uncertainty is larger, there are several differences between the two analyses.

In this work, we do not include SNe at $z>0.5$, and do not include the second and fourth moments.  Also, while the bootstrap re-sampling method we use here to estimate the data-covariance matrix allows us to estimate the covariance between the third moment and the correlation function, this method leads to larger uncertainties than the analytical covariance matrix used by \cite{2014MNRAS.443L...6C}.

Although the statistical uncertainty on the value of $\sigma_8$ is far from competitive with lensing measurements from galaxy shear techniques, we note the added-value nature of the measurement; placing limits on WL from observations taken to probe the distance-redshift relation.

We reiterate the conclusions from \cite{2017MNRAS.465.2862S} of the potential to place competitive constraints on WL with the next generations of supernovae surveys, such as LSST \citep{2009arXiv0912.0201L,2012arXiv1211.0310L,2019ApJ...873..111I}, Euclid \citep{2011arXiv1110.3193L,2014A&A...572A..80A,2018LRR....21....2A} and WFIRST \citep{2018ApJ...867...23H}.  We also emphasise the need for improved modelling of simulations and theory of the high redshift, small angular-scale lensing effects that will be required to fully realise the potential for SN-WL.

\section{Data Availability}

The supernova data underlying this article were accessed from \url{https://des.ncsa.illinois.edu/releases/sn}.  The simulated data were accessed from \url{http://maia.ice.cat/mice/}.

\section{Acknowledgements}

We thank the referee for a careful review, and constructive comments and suggestions, which have been very helpful in improving this work.

E.M., D.B., R.C.N. acknowledge funding from STFC grant ST/N000668/1.

This paper makes use of observations taken using the Anglo-Australian Telescope under programs ATAC A/2013B/12 and  NOAO 2013B-0317; the Gemini Observatory under programs NOAO 2013A-0373/GS-2013B-Q-45, NOAO 2015B-0197/GS-2015B-Q-7, and GS-2015B-Q-8; the Gran Telescopio Canarias under programs GTC77-13B, GTC70-14B, and GTC101-15B; the Keck Observatory under programs U063-2013B, U021-2014B, U048-2015B, U038-2016A; the Magellan Observatory under programs CN2015B-89; the MMT under 2014c-SAO-4, 2015a-SAO-12, 2015c-SAO-21; the South African Large Telescope under programs 2013-1-RSA\_OTH-023, 2013-2-RSA\_OTH-018, 2014-1-RSA\_OTH-016, 2014-2-SCI-070, 2015-1-SCI-063, and 2015-2-SCI-061; and the Very Large Telescope under programs ESO 093.A-0749(A), 094.A-0310(B), 095.A-0316(A), 096.A-0536(A), 095.D-0797(A). 

Funding for the DES Projects has been provided by the U.S. Department of Energy, the U.S. National Science Foundation, the Ministry of Science and Education of Spain, 
the Science and Technology Facilities Council of the United Kingdom, the Higher Education Funding Council for England, the National Center for Supercomputing 
Applications at the University of Illinois at Urbana-Champaign, the Kavli Institute of Cosmological Physics at the University of Chicago, 
the Center for Cosmology and Astro-Particle Physics at the Ohio State University,
the Mitchell Institute for Fundamental Physics and Astronomy at Texas A\&M University, Financiadora de Estudos e Projetos, 
Funda{\c c}{\~a}o Carlos Chagas Filho de Amparo {\`a} Pesquisa do Estado do Rio de Janeiro, Conselho Nacional de Desenvolvimento Cient{\'i}fico e Tecnol{\'o}gico and 
the Minist{\'e}rio da Ci{\^e}ncia, Tecnologia e Inova{\c c}{\~a}o, the Deutsche Forschungsgemeinschaft and the Collaborating Institutions in the Dark Energy Survey. 

The Collaborating Institutions are Argonne National Laboratory, the University of California at Santa Cruz, the University of Cambridge, Centro de Investigaciones Energ{\'e}ticas, 
Medioambientales y Tecnol{\'o}gicas-Madrid, the University of Chicago, University College London, the DES-Brazil Consortium, the University of Edinburgh, 
the Eidgen{\"o}ssische Technische Hochschule (ETH) Z{\"u}rich, 
Fermi National Accelerator Laboratory, the University of Illinois at Urbana-Champaign, the Institut de Ci{\`e}ncies de l'Espai (IEEC/CSIC), 
the Institut de F{\'i}sica d'Altes Energies, Lawrence Berkeley National Laboratory, the Ludwig-Maximilians Universit{\"a}t M{\"u}nchen and the associated Excellence Cluster Universe, 
the University of Michigan, the National Optical Astronomy Observatory, the University of Nottingham, The Ohio State University, the University of Pennsylvania, the University of Portsmouth, 
SLAC National Accelerator Laboratory, Stanford University, the University of Sussex, Texas A\&M University, and the OzDES Membership Consortium.

Based in part on observations at Cerro Tololo Inter-American Observatory, National Optical Astronomy Observatory, which is operated by the Association of Universities for Research in Astronomy (AURA) under a cooperative agreement with the National Science Foundation.

The DES data management system is supported by the National Science Foundation under Grant Numbers AST-1138766 and AST-1536171.
The DES participants from Spanish institutions are partially supported by MINECO under grants AYA2015-71825, ESP2015-66861, FPA2015-68048, SEV-2016-0588, SEV-2016-0597, and MDM-2015-0509, 
some of which include ERDF funds from the European Union. IFAE is partially funded by the CERCA program of the Generalitat de Catalunya.
Research leading to these results has received funding from the European Research
Council under the European Union's Seventh Framework Program (FP7/2007-2013) including ERC grant agreements 240672, 291329, and 306478.
We  acknowledge support from the Australian Research Council Centre of Excellence for All-sky Astrophysics (CAASTRO), through project number CE110001020, and the Brazilian Instituto Nacional de Ci\^encia
e Tecnologia (INCT) e-Universe (CNPq grant 465376/2014-2).

This manuscript has been authored by Fermi Research Alliance, LLC under Contract No. DE-AC02-07CH11359 with the U.S. Department of Energy, Office of Science, Office of High Energy Physics. The United States Government retains and the publisher, by accepting the article for publication, acknowledges that the United States Government retains a non-exclusive, paid-up, irrevocable, world-wide license to publish or reproduce the published form of this manuscript, or allow others to do so, for United States Government purposes.

\bibliographystyle{mnras}
\bibliography{references} 

\begin{thebibliography}{}
\makeatletter
\relax
\def\mn@urlcharsother{\let\do\@makeother \do\$\do\&\do\#\do\^\do\_\do\%\do\~}
\def\mn@doi{\begingroup\mn@urlcharsother \@ifnextchar [ {\mn@doi@}
  {\mn@doi@[]}}
\def\mn@doi@[#1]#2{\def\@tempa{#1}\ifx\@tempa\@empty \href
  {http://dx.doi.org/#2} {doi:#2}\else \href {http://dx.doi.org/#2} {#1}\fi
  \endgroup}
\def\mn@eprint#1#2{\mn@eprint@#1:#2::\@nil}
\def\mn@eprint@arXiv#1{\href {http://arxiv.org/abs/#1} {{\tt arXiv:#1}}}
\def\mn@eprint@dblp#1{\href {http://dblp.uni-trier.de/rec/bibtex/#1.xml}
  {dblp:#1}}
\def\mn@eprint@#1:#2:#3:#4\@nil{\def\@tempa {#1}\def\@tempb {#2}\def\@tempc
  {#3}\ifx \@tempc \@empty \let \@tempc \@tempb \let \@tempb \@tempa \fi \ifx
  \@tempb \@empty \def\@tempb {arXiv}\fi \@ifundefined
  {mn@eprint@\@tempb}{\@tempb:\@tempc}{\expandafter \expandafter \csname
  mn@eprint@\@tempb\endcsname \expandafter{\@tempc}}}

\bibitem[\protect\citeauthoryear{{Abbott} et~al.,}{{Abbott}
  et~al.}{2016}]{2016PhRvD..94b2001A}
{Abbott} T.,  et~al., 2016, \mn@doi [\prd] {10.1103/PhysRevD.94.022001}, \href
  {https://ui.adsabs.harvard.edu/abs/2016PhRvD..94b2001A} {94, 022001}

\bibitem[\protect\citeauthoryear{{Abbott} et~al.,}{{Abbott}
  et~al.}{2018}]{2018PhRvD..98d3526A}
{Abbott} T.~M.~C.,  et~al., 2018, \mn@doi [\prd] {10.1103/PhysRevD.98.043526},
  \href {https://ui.adsabs.harvard.edu/abs/2018PhRvD..98d3526A} {98, 043526}

\bibitem[\protect\citeauthoryear{{Abbott} et~al.,}{{Abbott}
  et~al.}{2019}]{2019ApJ...872L..30A}
{Abbott} T.~M.~C.,  et~al., 2019, \mn@doi [\apjl] {10.3847/2041-8213/ab04fa},
  \href {https://ui.adsabs.harvard.edu/abs/2019ApJ...872L..30A} {872, L30}

\bibitem[\protect\citeauthoryear{{Amendola} et~al.,}{{Amendola}
  et~al.}{2018}]{2018LRR....21....2A}
{Amendola} L.,  et~al., 2018, \mn@doi [Living Reviews in Relativity]
  {10.1007/s41114-017-0010-3}, \href
  {https://ui.adsabs.harvard.edu/abs/2018LRR....21....2A} {21, 2}

\bibitem[\protect\citeauthoryear{{Astier} et~al.,}{{Astier}
  et~al.}{2014}]{2014A&A...572A..80A}
{Astier} P.,  et~al., 2014, \mn@doi [\aap] {10.1051/0004-6361/201423551}, \href
  {https://ui.adsabs.harvard.edu/abs/2014A&A...572A..80A} {572, A80}

\bibitem[\protect\citeauthoryear{{Bartelmann} \& {Schneider}}{{Bartelmann} \&
  {Schneider}}{2001}]{2001PhR...340..291B}
{Bartelmann} M.,  {Schneider} P.,  2001, \mn@doi [\physrep]
  {10.1016/S0370-1573(00)00082-X}, \href
  {http://adsabs.harvard.edu/abs/2001PhR...340..291B} {340, 291}

\bibitem[\protect\citeauthoryear{{Bauer}, {Seitz}, {Jerke}, {Scalzo},
  {Rabinowitz}, {Ellman}  \& {Baltay}}{{Bauer}
  et~al.}{2011}]{2011ApJ...732...64B}
{Bauer} A.~H.,  {Seitz} S.,  {Jerke} J.,  {Scalzo} R.,  {Rabinowitz} D.,
  {Ellman} N.,   {Baltay} C.,  2011, \mn@doi [\apj]
  {10.1088/0004-637X/732/2/64}, \href
  {https://ui.adsabs.harvard.edu/abs/2011ApJ...732...64B} {732, 64}

\bibitem[\protect\citeauthoryear{{Baxter} et~al.,}{{Baxter}
  et~al.}{2019}]{2019PhRvD..99b3508B}
{Baxter} E.~J.,  et~al., 2019, \mn@doi [\prd] {10.1103/PhysRevD.99.023508},
  \href {https://ui.adsabs.harvard.edu/abs/2019PhRvD..99b3508B} {99, 023508}

\bibitem[\protect\citeauthoryear{{Becker} et~al.,}{{Becker}
  et~al.}{2016}]{2016PhRvD..94b2002B}
{Becker} M.~R.,  et~al., 2016, \mn@doi [\prd] {10.1103/PhysRevD.94.022002},
  \href {https://ui.adsabs.harvard.edu/abs/2016PhRvD..94b2002B} {94, 022002}

\bibitem[\protect\citeauthoryear{{Bertschinger} \& {Zukin}}{{Bertschinger} \&
  {Zukin}}{2008}]{2008PhRvD..78b4015B}
{Bertschinger} E.,  {Zukin} P.,  2008, \mn@doi [\prd]
  {10.1103/PhysRevD.78.024015}, \href
  {https://ui.adsabs.harvard.edu/abs/2008PhRvD..78b4015B} {78, 024015}

\bibitem[\protect\citeauthoryear{{Betoule} et~al.,}{{Betoule}
  et~al.}{2014}]{2014A&A...568A..22B}
{Betoule} M.,  et~al., 2014, \mn@doi [\aap] {10.1051/0004-6361/201423413},
  \href {http://adsabs.harvard.edu/abs/2014A%26A...568A..22B} {568, A22}

\bibitem[\protect\citeauthoryear{{Brout} et~al.,}{{Brout}
  et~al.}{2019a}]{2018arXiv181102378B}
{Brout} D.,  et~al., 2019a, \mn@doi [\apj] {10.3847/1538-4357/ab06c1}, \href
  {https://ui.adsabs.harvard.edu/abs/2019ApJ...874..106B} {874, 106}

\bibitem[\protect\citeauthoryear{{Brout} et~al.,}{{Brout}
  et~al.}{2019b}]{2018arXiv181102377B}
{Brout} D.,  et~al., 2019b, \mn@doi [\apj] {10.3847/1538-4357/ab08a0}, \href
  {https://ui.adsabs.harvard.edu/abs/2019ApJ...874..150B} {874, 150}

\bibitem[\protect\citeauthoryear{{Castro} \& {Quartin}}{{Castro} \&
  {Quartin}}{2014}]{2014MNRAS.443L...6C}
{Castro} T.,  {Quartin} M.,  2014, \mn@doi [\mnras] {10.1093/mnrasl/slu071},
  \href {https://ui.adsabs.harvard.edu/abs/2014MNRAS.443L...6C} {443, L6}

\bibitem[\protect\citeauthoryear{{Castro}, {Quartin}  \&
  {Benitez-Herrera}}{{Castro} et~al.}{2016}]{2016PDU....13...66C}
{Castro} T.,  {Quartin} M.,   {Benitez-Herrera} S.,  2016, \mn@doi [Physics of
  the Dark Universe] {10.1016/j.dark.2016.04.006}, \href
  {https://ui.adsabs.harvard.edu/abs/2016PDU....13...66C} {13, 66}

\bibitem[\protect\citeauthoryear{{Castro}, {Quartin}, {Giocoli}, {Borgani}  \&
  {Dolag}}{{Castro} et~al.}{2018}]{2018MNRAS.478.1305C}
{Castro} T.,  {Quartin} M.,  {Giocoli} C.,  {Borgani} S.,   {Dolag} K.,  2018,
  \mn@doi [\mnras] {10.1093/mnras/sty1117}, \href
  {https://ui.adsabs.harvard.edu/abs/2018MNRAS.478.1305C} {478, 1305}

\bibitem[\protect\citeauthoryear{{Challinor} \& {Lewis}}{{Challinor} \&
  {Lewis}}{2011}]{2011PhRvD..84d3516C}
{Challinor} A.,  {Lewis} A.,  2011, \mn@doi [\prd]
  {10.1103/PhysRevD.84.043516}, \href
  {https://ui.adsabs.harvard.edu/abs/2011PhRvD..84d3516C} {84, 043516}

\bibitem[\protect\citeauthoryear{{D'Andrea} et~al.,}{{D'Andrea}
  et~al.}{2018}]{2018arXiv181109565D}
{D'Andrea} C.~B.,  et~al., 2018, arXiv e-prints, \href
  {https://ui.adsabs.harvard.edu/\#abs/2018arXiv181109565D} {p.
  arXiv:1811.09565}

\bibitem[\protect\citeauthoryear{{Davis} et~al.,}{{Davis}
  et~al.}{2011}]{2011ApJ...741...67D}
{Davis} T.~M.,  et~al., 2011, \mn@doi [\apj] {10.1088/0004-637X/741/1/67},
  \href {https://ui.adsabs.harvard.edu/abs/2011ApJ...741...67D} {741, 67}

\bibitem[\protect\citeauthoryear{{Dawson} et~al.,}{{Dawson}
  et~al.}{2013}]{2013AJ....145...10D}
{Dawson} K.~S.,  et~al., 2013, \mn@doi [\aj] {10.1088/0004-6256/145/1/10},
  \href {https://ui.adsabs.harvard.edu/abs/2013AJ....145...10D} {145, 10}

\bibitem[\protect\citeauthoryear{{Dodelson} \& {Vallinotto}}{{Dodelson} \&
  {Vallinotto}}{2006}]{2006PhRvD..74f3515D}
{Dodelson} S.,  {Vallinotto} A.,  2006, \mn@doi [\prd]
  {10.1103/PhysRevD.74.063515}, \href
  {https://ui.adsabs.harvard.edu/abs/2006PhRvD..74f3515D} {74, 063515}

\bibitem[\protect\citeauthoryear{{Foreman-Mackey}, {Hogg}, {Lang}  \&
  {Goodman}}{{Foreman-Mackey} et~al.}{2013}]{2013PASP..125..306F}
{Foreman-Mackey} D.,  {Hogg} D.~W.,  {Lang} D.,   {Goodman} J.,  2013, \mn@doi
  [\pasp] {10.1086/670067}, \href
  {http://adsabs.harvard.edu/abs/2013PASP..125..306F} {125, 306}

\bibitem[\protect\citeauthoryear{{Fosalba}, {Gazta{\~n}aga}, {Castander}  \&
  {Crocce}}{{Fosalba} et~al.}{2015}]{2015MNRAS.447.1319F}
{Fosalba} P.,  {Gazta{\~n}aga} E.,  {Castander} F.~J.,   {Crocce} M.,  2015,
  \mn@doi [\mnras] {10.1093/mnras/stu2464}, \href
  {http://adsabs.harvard.edu/abs/2015MNRAS.447.1319F} {447, 1319}

\bibitem[\protect\citeauthoryear{{Garcia}, {Quartin}  \& {Siffert}}{{Garcia}
  et~al.}{2019}]{2019arXiv190500746G}
{Garcia} K.,  {Quartin} M.,   {Siffert} B.~B.,  2019, arXiv e-prints, \href
  {https://ui.adsabs.harvard.edu/abs/2019arXiv190500746G} {p. arXiv:1905.00746}

\bibitem[\protect\citeauthoryear{{Gordon}, {Land}  \& {Slosar}}{{Gordon}
  et~al.}{2007}]{2007PhRvL..99h1301G}
{Gordon} C.,  {Land} K.,   {Slosar} A.,  2007, \mn@doi [\prl]
  {10.1103/PhysRevLett.99.081301}, \href
  {https://ui.adsabs.harvard.edu/abs/2007PhRvL..99h1301G} {99, 081301}

\bibitem[\protect\citeauthoryear{{Heymans} et~al.,}{{Heymans}
  et~al.}{2012}]{2012MNRAS.427..146H}
{Heymans} C.,  et~al., 2012, \mn@doi [\mnras]
  {10.1111/j.1365-2966.2012.21952.x}, \href
  {https://ui.adsabs.harvard.edu/abs/2012MNRAS.427..146H} {427, 146}

\bibitem[\protect\citeauthoryear{{Hilbert}, {White}, {Hartlap}  \&
  {Schneider}}{{Hilbert} et~al.}{2008}]{2008MNRAS.386.1845H}
{Hilbert} S.,  {White} S. D.~M.,  {Hartlap} J.,   {Schneider} P.,  2008,
  \mn@doi [\mnras] {10.1111/j.1365-2966.2008.13190.x}, \href
  {https://ui.adsabs.harvard.edu/abs/2008MNRAS.386.1845H} {386, 1845}

\bibitem[\protect\citeauthoryear{{Hildebrandt} et~al.,}{{Hildebrandt}
  et~al.}{2017}]{2017MNRAS.465.1454H}
{Hildebrandt} H.,  et~al., 2017, \mn@doi [\mnras] {10.1093/mnras/stw2805},
  \href {http://adsabs.harvard.edu/abs/2017MNRAS.465.1454H} {465, 1454}

\bibitem[\protect\citeauthoryear{{Hounsell} et~al.,}{{Hounsell}
  et~al.}{2018}]{2018ApJ...867...23H}
{Hounsell} R.,  et~al., 2018, \mn@doi [\apj] {10.3847/1538-4357/aac08b}, \href
  {https://ui.adsabs.harvard.edu/abs/2018ApJ...867...23H} {867, 23}

\bibitem[\protect\citeauthoryear{{Hu} \& {Jain}}{{Hu} \&
  {Jain}}{2004}]{2004PhRvD..70d3009H}
{Hu} W.,  {Jain} B.,  2004, \mn@doi [\prd] {10.1103/PhysRevD.70.043009}, \href
  {https://ui.adsabs.harvard.edu/abs/2004PhRvD..70d3009H} {70, 043009}

\bibitem[\protect\citeauthoryear{{Ivezi{\'c}} et~al.,}{{Ivezi{\'c}}
  et~al.}{2019}]{2019ApJ...873..111I}
{Ivezi{\'c}} {\v{Z}}.,  et~al., 2019, \mn@doi [\apj]
  {10.3847/1538-4357/ab042c}, \href
  {https://ui.adsabs.harvard.edu/abs/2019ApJ...873..111I} {873, 111}

\bibitem[\protect\citeauthoryear{{Joudaki} et~al.,}{{Joudaki}
  et~al.}{2017}]{2017MNRAS.471.1259J}
{Joudaki} S.,  et~al., 2017, \mn@doi [\mnras] {10.1093/mnras/stx998}, \href
  {http://adsabs.harvard.edu/abs/2017MNRAS.471.1259J} {471, 1259}

\bibitem[\protect\citeauthoryear{{Kainulainen} \& {Marra}}{{Kainulainen} \&
  {Marra}}{2009}]{2009PhRvD..80l3020K}
{Kainulainen} K.,  {Marra} V.,  2009, \mn@doi [\prd]
  {10.1103/PhysRevD.80.123020}, \href
  {https://ui.adsabs.harvard.edu/abs/2009PhRvD..80l3020K} {80, 123020}

\bibitem[\protect\citeauthoryear{{Kainulainen} \& {Marra}}{{Kainulainen} \&
  {Marra}}{2011}]{2011PhRvD..83b3009K}
{Kainulainen} K.,  {Marra} V.,  2011, \mn@doi [\prd]
  {10.1103/PhysRevD.83.023009}, \href
  {https://ui.adsabs.harvard.edu/abs/2011PhRvD..83b3009K} {83, 023009}

\bibitem[\protect\citeauthoryear{{Kessler} et~al.,}{{Kessler}
  et~al.}{2019}]{2019MNRAS.tmp..472K}
{Kessler} R.,  et~al., 2019, \mn@doi [\mnras] {10.1093/mnras/stz463}, \href
  {https://ui.adsabs.harvard.edu/abs/2019MNRAS.485.1171K} {485, 1171}

\bibitem[\protect\citeauthoryear{{Knox}, {Song}  \& {Tyson}}{{Knox}
  et~al.}{2006}]{2006PhRvD..74b3512K}
{Knox} L.,  {Song} Y.-S.,   {Tyson} J.~A.,  2006, \mn@doi [\prd]
  {10.1103/PhysRevD.74.023512}, \href
  {https://ui.adsabs.harvard.edu/abs/2006PhRvD..74b3512K} {74, 023512}

\bibitem[\protect\citeauthoryear{{Kunz} \& {Sapone}}{{Kunz} \&
  {Sapone}}{2007}]{2007PhRvL..98l1301K}
{Kunz} M.,  {Sapone} D.,  2007, \mn@doi [\prl] {10.1103/PhysRevLett.98.121301},
  \href {https://ui.adsabs.harvard.edu/abs/2007PhRvL..98l1301K} {98, 121301}

\bibitem[\protect\citeauthoryear{{LSST Dark Energy Science
  Collaboration}}{{LSST Dark Energy Science
  Collaboration}}{2012}]{2012arXiv1211.0310L}
{LSST Dark Energy Science Collaboration} 2012, arXiv e-prints, \href
  {https://ui.adsabs.harvard.edu/abs/2012arXiv1211.0310L} {p. arXiv:1211.0310}

\bibitem[\protect\citeauthoryear{{LSST Science Collaboration} et~al.,}{{LSST
  Science Collaboration} et~al.}{2009}]{2009arXiv0912.0201L}
{LSST Science Collaboration} et~al., 2009, arXiv e-prints, \href
  {https://ui.adsabs.harvard.edu/abs/2009arXiv0912.0201L} {p. arXiv:0912.0201}

\bibitem[\protect\citeauthoryear{{Lasker} et~al.,}{{Lasker}
  et~al.}{2019}]{2018arXiv181102380L}
{Lasker} J.,  et~al., 2019, \mn@doi [\mnras] {10.1093/mnras/stz619}, \href
  {https://ui.adsabs.harvard.edu/abs/2019MNRAS.485.5329L} {485, 5329}

\bibitem[\protect\citeauthoryear{{Laureijs} et~al.,}{{Laureijs}
  et~al.}{2011}]{2011arXiv1110.3193L}
{Laureijs} R.,  et~al., 2011, arXiv e-prints, \href
  {https://ui.adsabs.harvard.edu/abs/2011arXiv1110.3193L} {p. arXiv:1110.3193}

\bibitem[\protect\citeauthoryear{{Liao}, {Fan}, {Ding}, {Biesiada}  \&
  {Zhu}}{{Liao} et~al.}{2017}]{2017NatCo...8.1148L}
{Liao} K.,  {Fan} X.-L.,  {Ding} X.,  {Biesiada} M.,   {Zhu} Z.-H.,  2017,
  \mn@doi [Nature Communications] {10.1038/s41467-017-01152-9}, \href
  {https://ui.adsabs.harvard.edu/abs/2017NatCo...8.1148L} {8, 1148}

\bibitem[\protect\citeauthoryear{{Macaulay}, {Davis}, {Scovacricchi}, {Bacon},
  {Collett}  \& {Nichol}}{{Macaulay} et~al.}{2017}]{2017MNRAS.467..259M}
{Macaulay} E.,  {Davis} T.~M.,  {Scovacricchi} D.,  {Bacon} D.,  {Collett} T.,
   {Nichol} R.~C.,  2017, \mn@doi [\mnras] {10.1093/mnras/stw3339}, \href
  {https://ui.adsabs.harvard.edu/abs/2017MNRAS.467..259M} {467, 259}

\bibitem[\protect\citeauthoryear{{Marra}, {Quartin}  \& {Amendola}}{{Marra}
  et~al.}{2013}]{2013PhRvD..88f3004M}
{Marra} V.,  {Quartin} M.,   {Amendola} L.,  2013, \mn@doi [\prd]
  {10.1103/PhysRevD.88.063004}, \href
  {http://adsabs.harvard.edu/abs/2013PhRvD..88f3004M} {88, 063004}

\bibitem[\protect\citeauthoryear{{Neill}, {Hudson}  \& {Conley}}{{Neill}
  et~al.}{2007}]{2007ApJ...661L.123N}
{Neill} J.~D.,  {Hudson} M.~J.,   {Conley} A.,  2007, \mn@doi [\apjl]
  {10.1086/518808}, \href
  {https://ui.adsabs.harvard.edu/abs/2007ApJ...661L.123N} {661, L123}

\bibitem[\protect\citeauthoryear{{Omori} et~al.,}{{Omori}
  et~al.}{2018}]{2018arXiv181002441O}
{Omori} Y.,  et~al., 2018, arXiv e-prints, \href
  {https://ui.adsabs.harvard.edu/abs/2018arXiv181002441O} {p. arXiv:1810.02441}

\bibitem[\protect\citeauthoryear{{Planck Collaboration} et~al.,}{{Planck
  Collaboration} et~al.}{2018a}]{2018arXiv180706209P}
{Planck Collaboration} et~al., 2018a, arXiv e-prints, \href
  {https://ui.adsabs.harvard.edu/abs/2018arXiv180706209P} {p. arXiv:1807.06209}

\bibitem[\protect\citeauthoryear{{Planck Collaboration} et~al.,}{{Planck
  Collaboration} et~al.}{2018b}]{2018arXiv180706210P}
{Planck Collaboration} et~al., 2018b, arXiv e-prints, \href
  {https://ui.adsabs.harvard.edu/abs/2018arXiv180706210P} {p. arXiv:1807.06210}

\bibitem[\protect\citeauthoryear{{Quartin}, {Marra}  \& {Amendola}}{{Quartin}
  et~al.}{2014}]{2014PhRvD..89b3009Q}
{Quartin} M.,  {Marra} V.,   {Amendola} L.,  2014, \mn@doi [\prd]
  {10.1103/PhysRevD.89.023009}, \href
  {http://adsabs.harvard.edu/abs/2014PhRvD..89b3009Q} {89, 023009}

\bibitem[\protect\citeauthoryear{{Schimd}, {Uzan}  \& {Riazuelo}}{{Schimd}
  et~al.}{2005}]{2005PhRvD..71h3512S}
{Schimd} C.,  {Uzan} J.-P.,   {Riazuelo} A.,  2005, \mn@doi [\prd]
  {10.1103/PhysRevD.71.083512}, \href
  {https://ui.adsabs.harvard.edu/abs/2005PhRvD..71h3512S} {71, 083512}

\bibitem[\protect\citeauthoryear{{Schimd} et~al.,}{{Schimd}
  et~al.}{2007}]{2007A&A...463..405S}
{Schimd} C.,  et~al., 2007, \mn@doi [\aap] {10.1051/0004-6361:20065154}, \href
  {https://ui.adsabs.harvard.edu/abs/2007A&A...463..405S} {463, 405}

\bibitem[\protect\citeauthoryear{{Schmidt}}{{Schmidt}}{2008}]{2008PhRvD..78d3002S}
{Schmidt} F.,  2008, \mn@doi [\prd] {10.1103/PhysRevD.78.043002}, \href
  {https://ui.adsabs.harvard.edu/abs/2008PhRvD..78d3002S} {78, 043002}

\bibitem[\protect\citeauthoryear{{Schrabback} et~al.,}{{Schrabback}
  et~al.}{2010}]{2010A&A...516A..63S}
{Schrabback} T.,  et~al., 2010, \mn@doi [\aap] {10.1051/0004-6361/200913577},
  \href {https://ui.adsabs.harvard.edu/abs/2010A&A...516A..63S} {516, A63}

\bibitem[\protect\citeauthoryear{{Scovacricchi}, {Nichol}, {Macaulay}  \&
  {Bacon}}{{Scovacricchi} et~al.}{2017}]{2017MNRAS.465.2862S}
{Scovacricchi} D.,  {Nichol} R.~C.,  {Macaulay} E.,   {Bacon} D.,  2017,
  \mn@doi [\mnras] {10.1093/mnras/stw2878}, \href
  {https://ui.adsabs.harvard.edu/abs/2017MNRAS.465.2862S} {465, 2862}

\bibitem[\protect\citeauthoryear{{Scranton} et~al.,}{{Scranton}
  et~al.}{2005}]{2005ApJ...633..589S}
{Scranton} R.,  et~al., 2005, \mn@doi [\apj] {10.1086/431358}, \href
  {https://ui.adsabs.harvard.edu/abs/2005ApJ...633..589S} {633, 589}

\bibitem[\protect\citeauthoryear{{Shang} \& {Haiman}}{{Shang} \&
  {Haiman}}{2011}]{2011MNRAS.411....9S}
{Shang} C.,  {Haiman} Z.,  2011, \mn@doi [\mnras]
  {10.1111/j.1365-2966.2010.17607.x}, \href
  {https://ui.adsabs.harvard.edu/abs/2011MNRAS.411....9S} {411, 9}

\bibitem[\protect\citeauthoryear{{Simpson} et~al.,}{{Simpson}
  et~al.}{2013}]{2013MNRAS.429.2249S}
{Simpson} F.,  et~al., 2013, \mn@doi [\mnras] {10.1093/mnras/sts493}, \href
  {https://ui.adsabs.harvard.edu/abs/2013MNRAS.429.2249S} {429, 2249}

\bibitem[\protect\citeauthoryear{Smith et~al.,}{Smith
  et~al.}{2003}]{Smith_2003}
Smith R.~E.,  et~al., 2003, \mn@doi [\mnras]
  {10.1046/j.1365-8711.2003.06503.x}, 341, 1311–1332

\bibitem[\protect\citeauthoryear{{Smith} et~al.,}{{Smith}
  et~al.}{2014}]{2014ApJ...780...24S}
{Smith} M.,  et~al., 2014, \mn@doi [\apj] {10.1088/0004-637X/780/1/24}, \href
  {http://adsabs.harvard.edu/abs/2014ApJ...780...24S} {780, 24}

\bibitem[\protect\citeauthoryear{{Springel}}{{Springel}}{2005}]{2005MNRAS.364.1105S}
{Springel} V.,  2005, \mn@doi [\mnras] {10.1111/j.1365-2966.2005.09655.x},
  \href {https://ui.adsabs.harvard.edu/abs/2005MNRAS.364.1105S} {364, 1105}

\bibitem[\protect\citeauthoryear{{Takahashi}, {Oguri}, {Sato}  \&
  {Hamana}}{{Takahashi} et~al.}{2011}]{2011ApJ...742...15T}
{Takahashi} R.,  {Oguri} M.,  {Sato} M.,   {Hamana} T.,  2011, \mn@doi [\apj]
  {10.1088/0004-637X/742/1/15}, \href
  {https://ui.adsabs.harvard.edu/abs/2011ApJ...742...15T} {742, 15}

\bibitem[\protect\citeauthoryear{{The Dark Energy Survey Collaboration}}{{The
  Dark Energy Survey Collaboration}}{2005}]{2005astro.ph.10346T}
{The Dark Energy Survey Collaboration} 2005, arXiv e-prints, \href
  {https://ui.adsabs.harvard.edu/abs/2005astro.ph.10346T} {pp
  astro--ph/0510346}

\bibitem[\protect\citeauthoryear{{Troxel} et~al.,}{{Troxel}
  et~al.}{2018}]{2018PhRvD..98d3528T}
{Troxel} M.~A.,  et~al., 2018, \mn@doi [\prd] {10.1103/PhysRevD.98.043528},
  \href {https://ui.adsabs.harvard.edu/abs/2018PhRvD..98d3528T} {98, 043528}

\bibitem[\protect\citeauthoryear{{Tsujikawa} \& {Tatekawa}}{{Tsujikawa} \&
  {Tatekawa}}{2008}]{2008PhLB..665..325T}
{Tsujikawa} S.,  {Tatekawa} T.,  2008, \mn@doi [Physics Letters B]
  {10.1016/j.physletb.2008.06.052}, \href
  {https://ui.adsabs.harvard.edu/abs/2008PhLB..665..325T} {665, 325}

\bibitem[\protect\citeauthoryear{{Valageas}}{{Valageas}}{2000}]{2000A&A...356..771V}
{Valageas} P.,  2000, \aap, \href
  {https://ui.adsabs.harvard.edu/abs/2000A&A...356..771V} {356, 771}

\bibitem[\protect\citeauthoryear{{Wambsganss}, {Cen}, {Xu}  \&
  {Ostriker}}{{Wambsganss} et~al.}{1997}]{1997ApJ...475L..81W}
{Wambsganss} J.,  {Cen} R.,  {Xu} G.,   {Ostriker} J.~P.,  1997, \mn@doi
  [\apjl] {10.1086/310470}, \href
  {https://ui.adsabs.harvard.edu/abs/1997ApJ...475L..81W} {475, L81}

\bibitem[\protect\citeauthoryear{{Wang}}{{Wang}}{2005}]{2005JCAP...03..005W}
{Wang} Y.,  2005, \mn@doi [\jcap] {10.1088/1475-7516/2005/03/005}, \href
  {https://ui.adsabs.harvard.edu/abs/2005JCAP...03..005W} {2005, 005}

\bibitem[\protect\citeauthoryear{{Wang} \& {Mukherjee}}{{Wang} \&
  {Mukherjee}}{2004}]{2004ApJ...606..654W}
{Wang} Y.,  {Mukherjee} P.,  2004, \mn@doi [\apj] {10.1086/383196}, \href
  {https://ui.adsabs.harvard.edu/abs/2004ApJ...606..654W} {606, 654}

\bibitem[\protect\citeauthoryear{{Wittman}, {Tyson}, {Kirkman}, {Dell'Antonio}
  \& {Bernstein}}{{Wittman} et~al.}{2000}]{2000Natur.405..143W}
{Wittman} D.~M.,  {Tyson} J.~A.,  {Kirkman} D.,  {Dell'Antonio} I.,
  {Bernstein} G.,  2000, \mn@doi [\nat] {10.1038/35012001}, \href
  {https://ui.adsabs.harvard.edu/abs/2000Natur.405..143W} {405, 143}

\bibitem[\protect\citeauthoryear{{Zhang}, {Liguori}, {Bean}  \&
  {Dodelson}}{{Zhang} et~al.}{2007}]{2007PhRvL..99n1302Z}
{Zhang} P.,  {Liguori} M.,  {Bean} R.,   {Dodelson} S.,  2007, \mn@doi [\prl]
  {10.1103/PhysRevLett.99.141302}, \href
  {https://ui.adsabs.harvard.edu/abs/2007PhRvL..99n1302Z} {99, 141302}

\bibitem[\protect\citeauthoryear{{Zuntz} et~al.,}{{Zuntz}
  et~al.}{2018}]{2018MNRAS.481.1149Z}
{Zuntz} J.,  et~al., 2018, \mn@doi [\mnras] {10.1093/mnras/sty2219}, \href
  {https://ui.adsabs.harvard.edu/abs/2018MNRAS.481.1149Z} {481, 1149}

\makeatother
\end{thebibliography}

\appendix

\section{Author Affiliations}
\label{appendix:affiliations}

$^{1}$ Department of Physics and Astronomy, University of North Georgia, Dahlonega, Georgia, 30597, USA\\
$^{2}$ Institute of Cosmology and Gravitation, University of Portsmouth, Portsmouth, PO1 3FX, UK\\
$^{3}$ School of Mathematics and Physics, University of Queensland,  Brisbane, QLD 4072, Australia\\
$^{4}$ Center for Cosmology and Astro-Particle Physics, The Ohio State University, Columbus, OH 43210, USA\\
$^{5}$ Department of Physics, The Ohio State University, Columbus, OH 43210, USA\\
$^{6}$ NASA Einstein Fellow\\
$^{7}$ Department of Physics and Astronomy, University of Pennsylvania, Philadelphia, PA 19104, USA\\
$^{8}$ INAF, Astrophysical Observatory of Turin, I-10025 Pino Torinese, Italy\\
$^{9}$ Centre for Astrophysics \& Supercomputing, Swinburne University of Technology, Victoria 3122, Australia\\
$^{10}$ Sydney Institute for Astronomy, School of Physics, A28, The University of Sydney, NSW 2006, Australia\\
$^{11}$ The Research School of Astronomy and Astrophysics, Australian National University, ACT 2601, Australia\\
$^{12}$ Universit\'e Clermont Auvergne, CNRS/IN2P3, LPC, F-63000 Clermont-Ferrand, France\\
$^{13}$ Department of Physics, Duke University Durham, NC 27708, USA\\
$^{14}$ School of Physics and Astronomy, University of Southampton,  Southampton, SO17 1BJ, UK\\
$^{15}$ Cerro Tololo Inter-American Observatory, National Optical Astronomy Observatory, Casilla 603, La Serena, Chile\\
$^{16}$ Departamento de F\'isica Matem\'atica, Instituto de F\'isica, Universidade de S\~ao Paulo, CP 66318, S\~ao Paulo, SP, 05314-970, Brazil\\
$^{17}$ Laborat\'orio Interinstitucional de e-Astronomia - LIneA, Rua Gal. Jos\'e Cristino 77, Rio de Janeiro, RJ - 20921-400, Brazil\\
$^{18}$ Fermi National Accelerator Laboratory, P. O. Box 500, Batavia, IL 60510, USA\\
$^{19}$ Instituto de Fisica Teorica UAM/CSIC, Universidad Autonoma de Madrid, 28049 Madrid, Spain\\
$^{20}$ CNRS, UMR 7095, Institut d'Astrophysique de Paris, F-75014, Paris, France\\
$^{21}$ Sorbonne Universit\'es, UPMC Univ Paris 06, UMR 7095, Institut d'Astrophysique de Paris, F-75014, Paris, France\\
$^{22}$ Department of Physics and Astronomy, Pevensey Building, University of Sussex, Brighton, BN1 9QH, UK\\
$^{23}$ Department of Physics \& Astronomy, University College London, Gower Street, London, WC1E 6BT, UK\\
$^{24}$ Kavli Institute for Particle Astrophysics \& Cosmology, P. O. Box 2450, Stanford University, Stanford, CA 94305, USA\\
$^{25}$ SLAC National Accelerator Laboratory, Menlo Park, CA 94025, USA\\
$^{26}$ Centro de Investigaciones Energ\'eticas, Medioambientales y Tecnol\'ogicas (CIEMAT), Madrid, Spain\\
$^{27}$ Department of Astronomy, University of Illinois at Urbana-Champaign, 1002 W. Green Street, Urbana, IL 61801, USA\\
$^{28}$ National Center for Supercomputing Applications, 1205 West Clark St., Urbana, IL 61801, USA\\
$^{29}$ Institut de F\'{\i}sica d'Altes Energies (IFAE), The Barcelona Institute of Science and Technology, Campus UAB, 08193 Bellaterra (Barcelona) Spain\\
$^{30}$ Institut d'Estudis Espacials de Catalunya (IEEC), 08034 Barcelona, Spain\\
$^{31}$ Institute of Space Sciences (ICE, CSIC),  Campus UAB, Carrer de Can Magrans, s/n,  08193 Barcelona, Spain\\
$^{32}$ INAF-Osservatorio Astronomico di Trieste, via G. B. Tiepolo 11, I-34143 Trieste, Italy\\
$^{33}$ Institute for Fundamental Physics of the Universe, Via Beirut 2, 34014 Trieste, Italy\\
$^{34}$ Observat\'orio Nacional, Rua Gal. Jos\'e Cristino 77, Rio de Janeiro, RJ - 20921-400, Brazil\\
$^{35}$ Department of Physics, IIT Hyderabad, Kandi, Telangana 502285, India\\
$^{36}$ Santa Cruz Institute for Particle Physics, Santa Cruz, CA 95064, USA\\
$^{37}$ Department of Astronomy, University of Michigan, Ann Arbor, MI 48109, USA\\
$^{38}$ Department of Physics, University of Michigan, Ann Arbor, MI 48109, USA\\
$^{39}$ Department of Physics, Stanford University, 382 Via Pueblo Mall, Stanford, CA 94305, USA\\
$^{40}$ Center for Astrophysics $\vert$ Harvard \& Smithsonian, 60 Garden Street, Cambridge, MA 02138, USA\\
$^{41}$ Australian Astronomical Optics, Macquarie University, North Ryde, NSW 2113, Australia\\
$^{42}$ Lowell Observatory, 1400 Mars Hill Rd, Flagstaff, AZ 86001, USA\\
$^{43}$ George P. and Cynthia Woods Mitchell Institute for Fundamental Physics and Astronomy, and Department of Physics and Astronomy, Texas A\&M University, College Station, TX 77843,  USA\\
$^{44}$ Department of Astrophysical Sciences, Princeton University, Peyton Hall, Princeton, NJ 08544, USA\\
$^{45}$ Instituci\'o Catalana de Recerca i Estudis Avan\c{c}ats, E-08010 Barcelona, Spain\\
$^{46}$ Kavli Institute for Cosmological Physics, University of Chicago, Chicago, IL 60637, USA\\
$^{47}$ Brandeis University, Physics Department, 415 South Street, Waltham MA 02453\\
$^{48}$ Computer Science and Mathematics Division, Oak Ridge National Laboratory, Oak Ridge, TN 37831\\
$^{49}$ Max Planck Institute for Extraterrestrial Physics, Giessenbachstrasse, 85748 Garching, Germany\\
$^{50}$ Universit\"ats-Sternwarte, Fakult\"at f\"ur Physik, Ludwig-Maximilians Universit\"at M\"unchen, Scheinerstr. 1, 81679 M\"unchen, Germany\\

\bsp	
\label{lastpage}
\end{document}